\documentclass[12pt]{iopart}
\usepackage{xcolor}
\usepackage{booktabs}

\input{definitions}

\let\Sectionmark\sectionmark
\def\sectionmark#1{\def\Sectionname{\uppercase{#1}}\Sectionmark{#1}}
\let\Subsectionmark\subsectionmark
\def\subsectionmark#1{\def\Subsectionname{#1}\Subsectionmark{#1}}

\DeclareAcronym{adv}{
  short=AdV,
  long=Advanced Virgo,
}

\DeclareAcronym{asd}{
  short=ASD,
  long=amplitude spectral density,
}

\DeclareAcronym{bh}{
  short=BH,
  long=black hole,
}

\DeclareAcronym{bns}{
  short=BNS,
  long=binary neutron star,
}

\DeclareAcronym{bristol}{
  short=\texttt{BRiSTOL},
  long=Band-limited \ac{rms} Stationarity Test Tool,
}

\DeclareAcronym{brms}{
  short=BRMS,
  long=Band-limited \ac{rms},
}

\DeclareAcronym{bruco}{
  short=\texttt{BruCo},
  long=brute-force coherence tool,
}

\DeclareAcronym{bs}{
  short=BS,
  long=beam splitter,
}

\DeclareAcronym{carm}{
  short=CARM,
  long=common (i.e. average) length of the two arm cavities,
}

\DeclareAcronym{ceb}{
  short=CEB,
  long=central building,
}

\DeclareAcronym{cw}{
  short=CW,
  long=continuous gravitational waves,
}

\DeclareAcronym{daq}{
  short=DAQ,
  long=data acquisition system,
}

\DeclareAcronym{darm}{
  short=DARM,
  long=difference of the two arm cavity lengths,
}

\DeclareAcronym{dms}{
  short=\texttt{DMS},
  long=Detector Monitoring System,
}

\DeclareAcronym{dof}{
  short=DOF,
  long=degree of freedom,
}

\DeclareAcronym{dq}{
  short=DQ,
  long=data quality,
}

\DeclareAcronym{dqr}{
  short=\texttt{DQR},
  long=Data Quality Report,
}

\DeclareAcronym{dqsegdb}{
  short=\texttt{DQSEGDB},
  long=Data Quality SEGment Database,
}

\DeclareAcronym{ego}{
  short=EGO,
  long=European Gravitational Observatory,
}

\DeclareAcronym{emd}{
  short=EMD,
  long=Empirical Mode Decomposition,
}

\DeclareAcronym{eom}{
  short=EOM,
  long=electro-optical modulator,
}

\DeclareAcronym{fft}{
  short=FFT,
  long=fast Fourier transform,
}

\DeclareAcronym{gracedb}{
  short=\texttt{GraceDB},
  long=GRAvitational-wave Candidate Event DataBase,
}

\DeclareAcronym{gw}{
  short=GW,
  long=gravitational wave,
}

\DeclareAcronym{gwosc}{
  short=GWOSC,
  long=Gravitational Wave Open Science Center,
}

\DeclareAcronym{imc}{
  short=IMC,
  long=input mode-cleaner,
}

\DeclareAcronym{kagra}{
  short=KAGRA,
  long=Kamioka Gravitational-Wave Detector,
}

\DeclareAcronym{ligo}{
  short=LIGO,
  long=Laser Interferometer Gravitational-wave Observatory,
}

\DeclareAcronym{lvalert}{
  short=\texttt{LVAlert},
  long=\ac{ligo}-Virgo Alert System,
}

\DeclareAcronym{mich}{
  short=MICH,
  long=length difference between the Virgo Michelson interferometer short arms
}

\DeclareAcronym{monet}{
  short=\texttt{MONET},
  long=Modulated NoisE Tool
}

\DeclareAcronym{ne}{
  short=NE,
  long=north end,
}

\DeclareAcronym{neb}{
  short=NEB,
  long=north-end building,
}

\DeclareAcronym{ni}{
  short=NI,
  long=north input,
}

\DeclareAcronym{noemi}{
  short=\texttt{NoEMi},
  long=Noise Frequency Event Miner,
}

\DeclareAcronym{ns}{
  short=NS,
  long=neutron star,
}

\DeclareAcronym{omc}{
  short=OMC,
  long=output mode-cleaner,
}

\DeclareAcronym{pr}{
  short=PR,
  long=power recycling,
}

\DeclareAcronym{prcl}{
  short=PRCL,
  long=power recycling cavity length,
}

\DeclareAcronym{psd}{
  short=PSD,
  long=power spectral density,
}

\DeclareAcronym{rms}{
  short=RMS,
  long=root mean square,
}

\DeclareAcronym{rrt}{
  short=RRT,
  long=rapid-response team,
}

\DeclareAcronym{sgwb}{
  short=SGWB,
  long=stochastic gravitational-wave background,
}

\DeclareAcronym{sneb}{
  short=SNEB,
  long=suspended north-end bench,
}

\DeclareAcronym{snr}{
  short=SNR,
  long=signal-to-noise ratio,
}

\DeclareAcronym{sr}{
  short=SR,
  long=signal recycling,
}

\DeclareAcronym{ssfs}{
  short=SSFS,
  long=second-stage frequency stabilization system,
}

\DeclareAcronym{sweb}{
  short=SWEB,
  long=suspended west-end bench,
}

\DeclareAcronym{upv}{
  short=\texttt{UPV},
  long=Use-Percentage Veto,
}

\DeclareAcronym{vim}{
  short=\texttt{VIM},
  long=Virgo Interferometer Monitor,
}

\DeclareAcronym{vpm}{
  short=VPM,
  long=Virgo Process Monitoring
}

\DeclareAcronym{we}{
  short=WE,
  long=west end,
}

\DeclareAcronym{web}{
  short=WEB,
  long=west-end building,
}

\DeclareAcronym{wi}{
  short=WI,
  long=west input,
}

\begin{document}
\leftline{Dated: \today}

\title{Virgo Detector Characterization and Data Quality: tools}

\author{%
F~Acernese$^{1,2}$, 
M~Agathos$^{3}$, 
A~Ain$^{4}$, 
S~Albanesi$^{5,6}$, 
A~Allocca\orcidlink{0000-0002-5288-1351}$^{7,2}$, 
A~Amato\orcidlink{0000-0001-9557-651X}$^{8}$, 
T~Andrade$^{9}$, 
N~Andres\orcidlink{0000-0002-5360-943X}$^{10}$, 
M~Andr\'es-Carcasona\orcidlink{0000-0002-8738-1672}$^{11}$, 
T~Andri\'c\orcidlink{0000-0002-9277-9773}$^{12}$, 
S~Ansoldi$^{13,14}$, 
S~Antier\orcidlink{0000-0002-7686-3334}$^{15,16}$, 
T~Apostolatos$^{17}$, 
E~Z~Appavuravther$^{18,19}$, %
M~Ar\`ene$^{20}$, %
N~Arnaud\orcidlink{0000-0001-6589-8673}$^{21,22}$, 
M~Assiduo$^{23,24}$, 
S~Assis~de~Souza~Melo$^{22}$, 
P~Astone\orcidlink{0000-0003-4981-4120}$^{25}$, 
F~Aubin\orcidlink{0000-0003-1613-3142}$^{24}$, 
S~Babak\orcidlink{0000-0001-7469-4250}$^{20}$, 
F~Badaracco\orcidlink{0000-0001-8553-7904}$^{26}$, 
M~K~M~Bader$^{27}$, %
S~Bagnasco\orcidlink{0000-0001-6062-6505}$^{6}$, 
J~Baird$^{20}$, %
T~Baka$^{28}$, 
G~Ballardin$^{22}$, 
G~Baltus\orcidlink{0000-0002-0304-8152}$^{29}$, 
B~Banerjee\orcidlink{0000-0002-8008-2485}$^{12}$, 
C~Barbieri$^{30,31,32}$, %
P~Barneo\orcidlink{0000-0002-8883-7280}$^{9}$, 
F~Barone\orcidlink{0000-0002-8069-8490}$^{33,2}$, 
M~Barsuglia\orcidlink{0000-0002-1180-4050}$^{20}$, 
D~Barta\orcidlink{0000-0001-6841-550X}$^{34}$, %
A~Basti$^{35,4}$, 
M~Bawaj\orcidlink{0000-0003-3611-3042}$^{18,36}$, 
M~Bazzan$^{37,38}$, 
F~Beirnaert\orcidlink{0000-0002-4003-7233}$^{39}$, 
M~Bejger\orcidlink{0000-0002-4991-8213}$^{40}$, 
I~Belahcene$^{21}$, %
V~Benedetto$^{41}$, %
M~Berbel\orcidlink{0000-0001-6345-1798}$^{42}$, 
S~Bernuzzi\orcidlink{0000-0002-2334-0935}$^{3}$, 
D~Bersanetti\orcidlink{0000-0002-7377-415X}$^{43}$, 
A~Bertolini$^{27}$, 
U~Bhardwaj\orcidlink{0000-0003-1233-4174}$^{16,27}$, 
A~Bianchi$^{27,44}$, 
S~Bini$^{45,46}$, 
M~Bischi$^{23,24}$, 
M~Bitossi$^{22,4}$, 
M-A~Bizouard\orcidlink{0000-0002-4618-1674}$^{15}$, 
F~Bobba$^{47,48}$, 
M~Bo\"{e}r$^{15}$, 
G~Bogaert$^{15}$, 
M~Boldrini$^{49,25}$, 
L~D~Bonavena$^{37}$, 
F~Bondu$^{50}$, 
R~Bonnand\orcidlink{0000-0001-5013-5913}$^{10}$, 
B~A~Boom$^{27}$, %
V~Boschi\orcidlink{0000-0001-8665-2293}$^{4}$, 
V~Boudart\orcidlink{0000-0001-9923-4154}$^{29}$, 
Y~Bouffanais$^{37,38}$, 
A~Bozzi$^{22}$, 
C~Bradaschia$^{4}$, 
M~Branchesi\orcidlink{0000-0003-1643-0526}$^{12,51}$, 
M~Breschi\orcidlink{0000-0002-3327-3676}$^{3}$, 
T~Briant\orcidlink{0000-0002-6013-1729}$^{52}$, 
A~Brillet$^{15}$, 
J~Brooks$^{22}$, %
G~Bruno$^{26}$, 
F~Bucci$^{24}$, 
T~Bulik$^{53}$, 
H~J~Bulten$^{27}$, 
D~Buskulic$^{10}$, 
C~Buy\orcidlink{0000-0003-2872-8186}$^{54}$, 
G~S~Cabourn~Davies\orcidlink{0000-0002-4289-3439}$^{55}$, %
G~Cabras\orcidlink{0000-0002-6852-6856}$^{13,14}$, 
R~Cabrita\orcidlink{0000-0003-0133-1306}$^{26}$, 
G~Cagnoli\orcidlink{0000-0002-7086-6550}$^{8}$, 
E~Calloni$^{7,2}$, 
M~Canepa$^{56,43}$, 
S~Canevarolo$^{28}$, 
M~Cannavacciuolo$^{47}$, %
E~Capocasa\orcidlink{0000-0003-3762-6958}$^{20}$, 
G~Carapella$^{47,48}$, 
F~Carbognani$^{22}$, 
M~Carpinelli$^{57,58,22}$, 
G~Carullo\orcidlink{0000-0001-9090-1862}$^{35,4}$, 
J~Casanueva~Diaz$^{22}$, 
C~Casentini$^{59,60}$, 
S~Caudill$^{27,28}$, 
F~Cavalier\orcidlink{0000-0002-3658-7240}$^{21}$, %
R~Cavalieri\orcidlink{0000-0001-6064-0569}$^{22}$, 
G~Cella\orcidlink{0000-0002-0752-0338}$^{4}$, 
P~Cerd\'a-Dur\'an$^{61}$, 
E~Cesarini\orcidlink{0000-0001-9127-3167}$^{60}$, 
W~Chaibi$^{15}$, 
P~Chanial\orcidlink{0000-0003-1753-524X}$^{22}$, %
E~Chassande-Mottin\orcidlink{0000-0003-3768-9908}$^{20}$, 
S~Chaty\orcidlink{0000-0002-5769-8601}$^{20}$, 
F~Chiadini\orcidlink{0000-0002-9339-8622}$^{62,48}$, 
G~Chiarini$^{38}$, 
R~Chierici$^{63}$, 
A~Chincarini\orcidlink{0000-0003-4094-9942}$^{43}$, 
M~L~Chiofalo$^{35,4}$, 
A~Chiummo\orcidlink{0000-0003-2165-2967}$^{22}$, 
S~Choudhary\orcidlink{0000-0003-0949-7298}$^{64}$, %
N~Christensen\orcidlink{0000-0002-6870-4202}$^{15}$, 
G~Ciani\orcidlink{0000-0003-4258-9338}$^{37,38}$, 
P~Ciecielag$^{40}$, 
M~Cie\'slar\orcidlink{0000-0001-8912-5587}$^{40}$, 
M~Cifaldi$^{59,60}$, 
R~Ciolfi\orcidlink{0000-0003-3140-8933}$^{65,38}$, 
F~Cipriano$^{15}$, %
S~Clesse$^{66}$, 
F~Cleva$^{15}$, 
E~Coccia$^{12,51}$, 
E~Codazzo\orcidlink{0000-0001-7170-8733}$^{12}$, 
P-F~Cohadon\orcidlink{0000-0003-3452-9415}$^{52}$, 
D~E~Cohen\orcidlink{0000-0002-0583-9919}$^{21}$, %
A~Colombo\orcidlink{0000-0002-7439-4773}$^{30,31}$, 
M~Colpi$^{30,31}$, 
L~Conti\orcidlink{0000-0003-2731-2656}$^{38}$, 
I~Cordero-Carri\'on\orcidlink{0000-0002-1985-1361}$^{67}$, 
S~Corezzi$^{36,18}$, 
D~Corre$^{21}$, %
S~Cortese\orcidlink{0000-0002-6504-0973}$^{22}$, 
J-P~Coulon$^{15}$, 
M~Croquette\orcidlink{0000-0002-8581-5393}$^{52}$, %
J~R~Cudell\orcidlink{0000-0002-2003-4238}$^{29}$, 
E~Cuoco$^{22,68,4}$, 
M~Cury{\l}o$^{53}$, 
P~Dabadie$^{8}$, %
T~Dal~Canton\orcidlink{0000-0001-5078-9044}$^{21}$, 
S~Dall'Osso\orcidlink{0000-0003-4366-8265}$^{12}$, %
G~D\'alya\orcidlink{0000-0003-3258-5763}$^{39}$, 
B~D'Angelo\orcidlink{0000-0001-9143-8427}$^{56,43}$, 
S~Danilishin\orcidlink{0000-0001-7758-7493}$^{69,27}$, 
S~D'Antonio$^{60}$, 
V~Dattilo$^{22}$, 
M~Davier$^{21}$, 
D~Davis\orcidlink{0000-0001-5620-6751}$^{70}$, %
J~Degallaix\orcidlink{0000-0002-1019-6911}$^{71}$, 
M~De~Laurentis$^{7,2}$, 
S~Del\'eglise\orcidlink{0000-0002-8680-5170}$^{52}$, 
F~De~Lillo\orcidlink{0000-0003-4977-0789}$^{26}$, 
D~Dell'Aquila\orcidlink{0000-0001-5895-0664}$^{57}$, 
W~Del~Pozzo$^{35,4}$, 
F~De~Matteis$^{59,60}$, %
A~Depasse\orcidlink{0000-0003-1014-8394}$^{26}$, 
R~De~Pietri\orcidlink{0000-0003-1556-8304}$^{72,73}$, 
R~De~Rosa\orcidlink{0000-0002-4004-947X}$^{7,2}$, 
C~De~Rossi$^{22}$, 
R~De~Simone$^{62}$, %
L~Di~Fiore$^{2}$, 
C~Di~Giorgio\orcidlink{0000-0003-2127-3991}$^{47,48}$, %
F~Di~Giovanni\orcidlink{0000-0001-8568-9334}$^{61}$, 
M~Di~Giovanni$^{12}$, 
T~Di~Girolamo\orcidlink{0000-0003-2339-4471}$^{7,2}$, 
A~Di~Lieto\orcidlink{0000-0002-4787-0754}$^{35,4}$, 
A~Di~Michele\orcidlink{0000-0002-0357-2608}$^{36}$, %
S~Di~Pace\orcidlink{0000-0001-6759-5676}$^{49,25}$, 
I~Di~Palma\orcidlink{0000-0003-1544-8943}$^{49,25}$, 
F~Di~Renzo\orcidlink{0000-0002-5447-3810}$^{35,4}$, 
L~D'Onofrio\orcidlink{0000-0001-9546-5959}$^{7,2}$, %
M~Drago\orcidlink{0000-0002-3738-2431}$^{49,25}$, 
J-G~Ducoin$^{21}$, 
U~Dupletsa$^{12}$, 
O~Durante$^{47,48}$, 
D~D'Urso\orcidlink{0000-0002-8215-4542}$^{57,58}$, 
P-A~Duverne$^{21}$, 
M~Eisenmann$^{10}$, %
L~Errico$^{7,2}$, 
D~Estevez\orcidlink{0000-0002-3021-5964}$^{74}$, 
F~Fabrizi\orcidlink{0000-0002-3809-065X}$^{23,24}$, 
F~Faedi$^{24}$, 
V~Fafone\orcidlink{0000-0003-1314-1622}$^{59,60,12}$, 
S~Farinon$^{43}$, %
G~Favaro\orcidlink{0000-0002-0351-6833}$^{37}$, 
M~Fays\orcidlink{0000-0002-4390-9746}$^{29}$, 
E~Fenyvesi\orcidlink{0000-0003-2777-3719}$^{34,75}$, %
I~Ferrante\orcidlink{0000-0002-0083-7228}$^{35,4}$, 
F~Fidecaro\orcidlink{0000-0002-6189-3311}$^{35,4}$, 
P~Figura\orcidlink{0000-0002-8925-0393}$^{53}$, 
A~Fiori\orcidlink{0000-0003-3174-0688}$^{4,35}$, 
I~Fiori\orcidlink{0000-0002-0210-516X}$^{22}$, 
R~Fittipaldi$^{76,48}$, %
V~Fiumara$^{77,48}$, %
R~Flaminio$^{10,78}$, 
J~A~Font\orcidlink{0000-0001-6650-2634}$^{61,79}$, 
S~Frasca$^{49,25}$, 
F~Frasconi\orcidlink{0000-0003-4204-6587}$^{4}$, 
A~Freise\orcidlink{0000-0001-6586-9901}$^{27,44}$, 
O~Freitas$^{80}$, 
G~G~Fronz\'e\orcidlink{0000-0003-0966-4279}$^{6}$, 
B~U~Gadre\orcidlink{0000-0002-1534-9761}$^{81,28}$, %
R~Gamba$^{3}$, 
B~Garaventa\orcidlink{0000-0003-2490-404X}$^{43,56}$, 
F~Garufi\orcidlink{0000-0003-1391-6168}$^{7,2}$, 
G~Gemme\orcidlink{0000-0002-1127-7406}$^{43}$, 
A~Gennai\orcidlink{0000-0003-0149-2089}$^{4}$, 
Archisman~Ghosh\orcidlink{0000-0003-0423-3533}$^{39}$, 
B~Giacomazzo\orcidlink{0000-0002-6947-4023}$^{30,31,32}$, 
L~Giacoppo$^{49,25}$, %
P~Giri\orcidlink{0000-0002-4628-2432}$^{4,35}$, 
F~Gissi$^{41}$, %
S~Gkaitatzis\orcidlink{0000-0001-9420-7499}$^{4,35}$, %
B~Goncharov\orcidlink{0000-0003-3189-5807}$^{12}$, 
M~Gosselin$^{22}$, 
R~Gouaty$^{10}$, 
A~Grado\orcidlink{0000-0002-0501-8256}$^{82,2}$, 
M~Granata\orcidlink{0000-0003-3275-1186}$^{71}$, 
V~Granata$^{47}$, 
G~Greco$^{18}$, 
G~Grignani$^{36,18}$, 
A~Grimaldi\orcidlink{0000-0002-6956-4301}$^{45,46}$, 
S~J~Grimm$^{12,51}$, %
P~Gruning$^{21}$, %
D~Guerra\orcidlink{0000-0003-0029-5390}$^{61}$, 
G~M~Guidi\orcidlink{0000-0002-3061-9870}$^{23,24}$, 
G~Guix\'e$^{9}$, 
Y~Guo$^{27}$, 
P~Gupta$^{27,28}$, 
L~Haegel\orcidlink{0000-0002-3680-5519}$^{20}$, 
O~Halim\orcidlink{0000-0003-1326-5481}$^{14}$, 
O~Hannuksela$^{28,27}$, 
T~Harder$^{15}$, 
K~Haris$^{27,28}$, 
J~Harms\orcidlink{0000-0002-7332-9806}$^{12,51}$, 
B~Haskell$^{40}$, 
A~Heidmann\orcidlink{0000-0002-0784-5175}$^{52}$, 
H~Heitmann\orcidlink{0000-0003-0625-5461}$^{15}$, 
P~Hello$^{21}$, 
G~Hemming\orcidlink{0000-0001-5268-4465}$^{22}$, 
E~Hennes\orcidlink{0000-0002-2246-5496}$^{27}$, 
S~Hild$^{69,27}$, 
D~Hofman$^{71}$, %
V~Hui\orcidlink{0000-0002-0233-2346}$^{10}$, 
B~Idzkowski\orcidlink{0000-0001-5869-2714}$^{53}$, 
A~Iess$^{59,60}$, %
P~Iosif\orcidlink{0000-0003-1621-7709}$^{83}$, 
T~Jacqmin\orcidlink{0000-0002-0693-4838}$^{52}$, 
P-E~Jacquet\orcidlink{0000-0001-9552-0057}$^{52}$, %
S~P~Jadhav$^{64}$, %
J~Janquart$^{28,27}$, 
K~Janssens\orcidlink{0000-0001-8760-4429}$^{84,15}$, 
P~Jaranowski\orcidlink{0000-0001-8085-3414}$^{85}$, 
V~Juste$^{74}$, 
C~Kalaghatgi$^{28,27,86}$, 
C~Karathanasis\orcidlink{0000-0002-0642-5507}$^{11}$, 
S~Katsanevas\orcidlink{0000-0003-0324-0758}$^{22}$\footnote{Deceased, November 2022.}, 
F~K\'ef\'elian$^{15}$, 
N~Khetan$^{12,51}$, %
G~Koekoek$^{27,69}$, 
S~Koley\orcidlink{0000-0002-5793-6665}$^{12}$, 
M~Kolstein\orcidlink{0000-0002-5482-6743}$^{11}$, 
A~Kr\'olak\orcidlink{0000-0003-4514-7690}$^{87,88}$, 
P~Kuijer\orcidlink{0000-0002-6987-2048}$^{27}$, 
P~Lagabbe$^{10}$, 
D~Laghi\orcidlink{0000-0001-7462-3794}$^{54}$, 
M~Lalleman$^{84}$, 
A~Lamberts$^{15,89}$, 
I~La~Rosa$^{10}$, 
A~Lartaux-Vollard$^{21}$, %
C~Lazzaro$^{37,38}$, 
P~Leaci\orcidlink{0000-0002-3997-5046}$^{49,25}$, 
A~Lema{\^i}tre$^{90}$, 
M~Lenti\orcidlink{0000-0002-2765-3955}$^{24,91}$, 
E~Leonova$^{16}$, %
N~Leroy\orcidlink{0000-0002-2321-1017}$^{21}$, 
N~Letendre$^{10}$, 
K~Leyde$^{20}$, 
F~Linde$^{86,27}$, 
L~London$^{16}$, 
A~Longo\orcidlink{0000-0003-4254-8579}$^{92}$, 
M~Lopez~Portilla$^{28}$, %
M~Lorenzini\orcidlink{0000-0002-2765-7905}$^{59,60}$, 
V~Loriette$^{93}$, 
G~Losurdo\orcidlink{0000-0003-0452-746X}$^{4}$, 
D~Lumaca\orcidlink{0000-0002-3628-1591}$^{59,60}$, 
A~Macquet$^{15}$, 
C~Magazz\`u\orcidlink{0000-0002-9913-381X}$^{4}$, %
M~Magnozzi\orcidlink{0000-0003-4512-8430}$^{43,56}$, 
E~Majorana$^{49,25}$, 
I~Maksimovic$^{93}$, %
N~Man$^{15}$, 
V~Mangano\orcidlink{0000-0001-7902-8505}$^{49,25}$, 
M~Mantovani\orcidlink{0000-0002-4424-5726}$^{22}$, 
M~Mapelli\orcidlink{0000-0001-8799-2548}$^{37,38}$, 
F~Marchesoni$^{19,18,94}$, 
D~Mar\'{\i}n~Pina\orcidlink{0000-0001-6482-1842}$^{9}$, 
F~Marion$^{10}$, 
A~Marquina$^{67}$, 
S~Marsat\orcidlink{0000-0001-9449-1071}$^{20}$, 
F~Martelli$^{23,24}$, 
M~Martinez$^{11}$, 
V~Martinez$^{8}$, 
A~Masserot$^{10}$, 
S~Mastrogiovanni\orcidlink{0000-0003-1606-4183}$^{20}$, 
Q~Meijer$^{28}$, 
A~Menendez-Vazquez$^{11}$, %
L~Mereni$^{71}$, 
M~Merzougui$^{15}$, %
A~Miani\orcidlink{0000-0001-7737-3129}$^{45,46}$, 
C~Michel\orcidlink{0000-0003-0606-725X}$^{71}$, 
L~Milano$^{7}$\footnote{Deceased, April 2021.}, %
A~Miller$^{26}$, 
B~Miller$^{16,27}$, %
E~Milotti$^{95,14}$, 
Y~Minenkov$^{60}$, 
Ll~M~Mir$^{11}$, 
M~Miravet-Ten\'es\orcidlink{0000-0002-8766-1156}$^{61}$, 
M~Montani$^{23,24}$, 
F~Morawski$^{40}$, 
B~Mours\orcidlink{0000-0002-6444-6402}$^{74}$, 
C~M~Mow-Lowry\orcidlink{0000-0002-0351-4555}$^{27,44}$
% ...
}%
\address{$^{1}$Dipartimento di Farmacia, Universit\`a di Salerno, I-84084 Fisciano, Salerno, Italy}
\address{$^{2}$INFN, Sezione di Napoli, Complesso Universitario di Monte S. Angelo, I-80126 Napoli, Italy}
\address{$^{3}$Theoretisch-Physikalisches Institut, Friedrich-Schiller-Universit\"at Jena, D-07743 Jena, Germany}
\address{$^{4}$INFN, Sezione di Pisa, I-56127 Pisa, Italy}
\address{$^{5}$Dipartimento di Fisica, Universit\`a degli Studi di Torino, I-10125 Torino, Italy}
\address{$^{6}$INFN Sezione di Torino, I-10125 Torino, Italy}
\address{$^{7}$Universit\`a di Napoli ``Federico II'', Complesso Universitario di Monte S. Angelo, I-80126 Napoli, Italy}
\address{$^{8}$Universit\'e de Lyon, Universit\'e Claude Bernard Lyon 1, CNRS, Institut Lumi\`ere Mati\`ere, F-69622 Villeurbanne, France}
\address{$^{9}$Institut de Ci\`encies del Cosmos (ICCUB), Universitat de Barcelona, C/ Mart\'{\i} i Franqu\`es 1, Barcelona, 08028, Spain}
\address{$^{10}$Univ. Savoie Mont Blanc, CNRS, Laboratoire d'Annecy de Physique des Particules - IN2P3, F-74000 Annecy, France}
\address{$^{11}$Institut de F\'{\i}sica d'Altes Energies (IFAE), Barcelona Institute of Science and Technology, and  ICREA, E-08193 Barcelona, Spain}
\address{$^{12}$Gran Sasso Science Institute (GSSI), I-67100 L'Aquila, Italy}
\address{$^{13}$Dipartimento di Scienze Matematiche, Informatiche e Fisiche, Universit\`a di Udine, I-33100 Udine, Italy}
\address{$^{14}$INFN, Sezione di Trieste, I-34127 Trieste, Italy}
\address{$^{15}$Artemis, Universit\'e C\^ote d'Azur, Observatoire de la C\^ote d'Azur, CNRS, F-06304 Nice, France}
\address{$^{16}$GRAPPA, Anton Pannekoek Institute for Astronomy and Institute for High-Energy Physics, University of Amsterdam, Science Park 904, 1098 XH Amsterdam, Netherlands}
\address{$^{17}$Department of Physics, National and Kapodistrian University of Athens, School of Science Building, 2nd floor, Panepistimiopolis, 15771 Ilissia, Greece}
\address{$^{18}$INFN, Sezione di Perugia, I-06123 Perugia, Italy}
\address{$^{19}$Universit\`a di Camerino, I-62032 Camerino, Italy}
\address{$^{20}$Universit\'e de Paris, CNRS, Astroparticule et Cosmologie, F-75006 Paris, France}
\address{$^{21}$Universit\'e Paris-Saclay, CNRS/IN2P3, IJCLab, 91405 Orsay, France}
\address{$^{22}$European Gravitational Observatory (EGO), I-56021 Cascina, Pisa, Italy}
\address{$^{23}$Universit\`a degli Studi di Urbino ``Carlo Bo'', I-61029 Urbino, Italy}
\address{$^{24}$INFN, Sezione di Firenze, I-50019 Sesto Fiorentino, Firenze, Italy}
\address{$^{25}$INFN, Sezione di Roma, I-00185 Roma, Italy}
\address{$^{26}$Universit\'e catholique de Louvain, B-1348 Louvain-la-Neuve, Belgium}
\address{$^{27}$Nikhef, Science Park 105, 1098 XG Amsterdam, Netherlands}
\address{$^{28}$Institute for Gravitational and Subatomic Physics (GRASP), Utrecht University, Princetonplein 1, 3584 CC Utrecht, Netherlands}
\address{$^{29}$Universit\'e de Li\`ege, B-4000 Li\`ege, Belgium}
\address{$^{30}$Universit\`a degli Studi di Milano-Bicocca, I-20126 Milano, Italy}
\address{$^{31}$INFN, Sezione di Milano-Bicocca, I-20126 Milano, Italy}
\address{$^{32}$INAF, Osservatorio Astronomico di Brera sede di Merate, I-23807 Merate, Lecco, Italy}
\address{$^{33}$Dipartimento di Medicina, Chirurgia e Odontoiatria ``Scuola Medica Salernitana'', Universit\`a di Salerno, I-84081 Baronissi, Salerno, Italy}
\address{$^{34}$Wigner RCP, RMKI, H-1121 Budapest, Konkoly Thege Mikl\'os \'ut 29-33, Hungary}
\address{$^{35}$Universit\`a di Pisa, I-56127 Pisa, Italy}
\address{$^{36}$Universit\`a di Perugia, I-06123 Perugia, Italy}
\address{$^{37}$Universit\`a di Padova, Dipartimento di Fisica e Astronomia, I-35131 Padova, Italy}
\address{$^{38}$INFN, Sezione di Padova, I-35131 Padova, Italy}
\address{$^{39}$Universiteit Gent, B-9000 Gent, Belgium}
\address{$^{40}$Nicolaus Copernicus Astronomical Center, Polish Academy of Sciences, 00-716, Warsaw, Poland}
\address{$^{41}$Dipartimento di Ingegneria, Universit\`a del Sannio, I-82100 Benevento, Italy}
\address{$^{42}$Departamento de Matem\'aticas, Universitat Aut\`onoma de Barcelona, Edificio C Facultad de Ciencias 08193 Bellaterra (Barcelona), Spain}
\address{$^{43}$INFN, Sezione di Genova, I-16146 Genova, Italy}
\address{$^{44}$Vrije Universiteit Amsterdam, 1081 HV Amsterdam, Netherlands}
\address{$^{45}$Universit\`a di Trento, Dipartimento di Fisica, I-38123 Povo, Trento, Italy}
\address{$^{46}$INFN, Trento Institute for Fundamental Physics and Applications, I-38123 Povo, Trento, Italy}
\address{$^{47}$Dipartimento di Fisica ``E.R. Caianiello'', Universit\`a di Salerno, I-84084 Fisciano, Salerno, Italy}
\address{$^{48}$INFN, Sezione di Napoli, Gruppo Collegato di Salerno, Complesso Universitario di Monte S. Angelo, I-80126 Napoli, Italy}
\address{$^{49}$Universit\`a di Roma ``La Sapienza'', I-00185 Roma, Italy}
\address{$^{50}$Univ Rennes, CNRS, Institut FOTON - UMR6082, F-3500 Rennes, France}
\address{$^{51}$INFN, Laboratori Nazionali del Gran Sasso, I-67100 Assergi, Italy}
\address{$^{52}$Laboratoire Kastler Brossel, Sorbonne Universit\'e, CNRS, ENS-Universit\'e PSL, Coll\`ege de France, F-75005 Paris, France}
\address{$^{53}$Astronomical Observatory Warsaw University, 00-478 Warsaw, Poland}
\address{$^{54}$L2IT, Laboratoire des 2 Infinis - Toulouse, Universit\'e de Toulouse, CNRS/IN2P3, UPS, F-31062 Toulouse Cedex 9, France}
\address{$^{55}$University of Portsmouth, Portsmouth, PO1 3FX, United Kingdom}
\address{$^{56}$Dipartimento di Fisica, Universit\`a degli Studi di Genova, I-16146 Genova, Italy}
\address{$^{57}$Universit\`a degli Studi di Sassari, I-07100 Sassari, Italy}
\address{$^{58}$INFN, Laboratori Nazionali del Sud, I-95125 Catania, Italy}
\address{$^{59}$Universit\`a di Roma Tor Vergata, I-00133 Roma, Italy}
\address{$^{60}$INFN, Sezione di Roma Tor Vergata, I-00133 Roma, Italy}
\address{$^{61}$Departamento de Astronom\'{\i}a y Astrof\'{\i}sica, Universitat de Val\`encia, E-46100 Burjassot, Val\`encia, Spain}
\address{$^{62}$Dipartimento di Ingegneria Industriale (DIIN), Universit\`a di Salerno, I-84084 Fisciano, Salerno, Italy}
\address{$^{63}$Universit\'e Lyon, Universit\'e Claude Bernard Lyon 1, CNRS, IP2I Lyon / IN2P3, UMR 5822, F-69622 Villeurbanne, France}
\address{$^{64}$Inter-University Centre for Astronomy and Astrophysics, Post Bag 4, Ganeshkhind, Pune 411 007, India}
\address{$^{65}$INAF, Osservatorio Astronomico di Padova, I-35122 Padova, Italy}
\address{$^{66}$Universit\'e libre de Bruxelles, Avenue Franklin Roosevelt 50 - 1050 Bruxelles, Belgium}
\address{$^{67}$Departamento de Matem\'aticas, Universitat de Val\`encia, E-46100 Burjassot, Val\`encia, Spain}
\address{$^{68}$Scuola Normale Superiore, Piazza dei Cavalieri, 7 - 56126 Pisa, Italy}
\address{$^{69}$Maastricht University, P.O. Box 616, 6200 MD Maastricht, Netherlands}
\address{$^{70}$LIGO Laboratory, California Institute of Technology, Pasadena, CA 91125, USA}
\address{$^{71}$Universit\'e Lyon, Universit\'e Claude Bernard Lyon 1, CNRS, Laboratoire des Mat\'eriaux Avanc\'es (LMA), IP2I Lyon / IN2P3, UMR 5822, F-69622 Villeurbanne, France}
\address{$^{72}$Dipartimento di Scienze Matematiche, Fisiche e Informatiche, Universit\`a di Parma, I-43124 Parma, Italy}
\address{$^{73}$INFN, Sezione di Milano Bicocca, Gruppo Collegato di Parma, I-43124 Parma, Italy}
\address{$^{74}$Universit\'e de Strasbourg, CNRS, IPHC UMR 7178, F-67000 Strasbourg, France}
\address{$^{75}$Institute for Nuclear Research, Bem t'er 18/c, H-4026 Debrecen, Hungary}
\address{$^{76}$CNR-SPIN, c/o Universit\`a di Salerno, I-84084 Fisciano, Salerno, Italy}
\address{$^{77}$Scuola di Ingegneria, Universit\`a della Basilicata, I-85100 Potenza, Italy}
\address{$^{78}$Gravitational Wave Science Project, National Astronomical Observatory of Japan (NAOJ), Mitaka City, Tokyo 181-8588, Japan}
\address{$^{79}$Observatori Astron\`omic, Universitat de Val\`encia, E-46980 Paterna, Val\`encia, Spain}
\address{$^{80}$Centro de F\'{\i}sica das Universidades do Minho e do Porto, Universidade do Minho, Campus de Gualtar, PT-4710 - 057 Braga, Portugal}
\address{$^{81}$Max Planck Institute for Gravitational Physics (Albert Einstein Institute), D-14476 Potsdam, Germany}
\address{$^{82}$INAF, Osservatorio Astronomico di Capodimonte, I-80131 Napoli, Italy}
\address{$^{83}$Department of Physics, Aristotle University of Thessaloniki, University Campus, 54124 Thessaloniki, Greece}
\address{$^{84}$Universiteit Antwerpen, Prinsstraat 13, 2000 Antwerpen, Belgium}
\address{$^{85}$University of Bia{\l}ystok, 15-424 Bia{\l}ystok, Poland}
\address{$^{86}$Institute for High-Energy Physics, University of Amsterdam, Science Park 904, 1098 XH Amsterdam, Netherlands}
\address{$^{87}$Institute of Mathematics, Polish Academy of Sciences, 00656 Warsaw, Poland}
\address{$^{88}$National Center for Nuclear Research, 05-400 {\' S}wierk-Otwock, Poland}
\address{$^{89}$Laboratoire Lagrange, Universit\'e C\^ote d'Azur, Observatoire C\^ote d'Azur, CNRS, F-06304 Nice, France}
\address{$^{90}$NAVIER, \'{E}cole des Ponts, Univ Gustave Eiffel, CNRS, Marne-la-Vall\'{e}e, France}
\address{$^{91}$Universit\`a di Firenze, Sesto Fiorentino I-50019, Italy}
\address{$^{92}$INFN, Sezione di Roma Tre, I-00146 Roma, Italy}
\address{$^{93}$ESPCI, CNRS, F-75005 Paris, France}
\address{$^{94}$School of Physics Science and Engineering, Tongji University, Shanghai 200092, China}
\address{$^{95}$Dipartimento di Fisica, Universit\`a di Trieste, I-34127 Trieste, Italy}
% ...

\begin{abstract}
Detector characterization and data quality studies --- collectively referred to as
{\em DetChar} activities in this article --- are paramount to the scientific
exploitation of the joint dataset collected by the LIGO-Virgo-KAGRA global network
of ground-based gravitational-wave (GW) detectors. They take place during each
phase of the operation of the instruments (upgrade, tuning and optimization,
data taking), are required at all steps of the dataflow (from data acquisition
to the final list of GW events) and operate at various latencies (from near real-time to
vet the public alerts to offline analyses).
This work requires a wide set of tools which have been developed over the years
to fulfill the requirements of the various DetChar studies: data access and
bookkeeping; global monitoring of the instruments and of the different steps of
the data processing; studies of the global properties of the noise at the 
detector outputs; identification and follow-up of noise peculiar features
(whether they be transient or continuously present in the data); quick processing
of the public alerts.
The present article reviews all the tools used by the Virgo DetChar
group during the third LIGO-Virgo Observation Run (O3, from April 2019 to March 2020),
mainly to analyse the Virgo data acquired at EGO.
Concurrently, a companion article focuses on the results achieved
by the DetChar group during the O3 run using these tools.

\end{abstract}

\maketitle

\tableofcontents

\clearpage

\mainmatter

\setlength{\parindent}{0pt}
\setlength{\parskip}{\medskipamount}


\section{Introduction}
\markboth{\thesection. \Sectionname}{}

\subsection{Background: DetChar inputs to detect and study gravitational waves}

GWTC-3~\cite{GWTC3}, the most recent edition of the \ac{gw} Transient Catalog edited by the \ac{ligo}~\cite{TheLIGOScientific:2014jea}, Virgo~\cite{TheVirgo:2014hva}, and now \ac{kagra}~\cite{10.1093/ptep/ptab018} scientific collaborations includes 90 \ac{gw} signals recorded between 2015 and 2020 during three successive data-taking campaigns called Observation Runs (in short O$_n$ with $n=1,2,3$). These discoveries are the result of the joint work of hundreds of scientists worldwide, bringing together a wide range of expertise, ranging from the instrumental side to data analyses. Important components of this global effort are the DetChar activities~\cite{Aasi:2012wd,LIGOScientific:2014qfs} that focus on studying the detector noises in all their variety: detector characterization on the one hand and data quality studies on the other.

Indeed, the Virgo \ac{gw} strain channel $h(t)$, reconstructed from its raw data, is dominated by noises of various origins (fundamental, technical or environmental~\cite{o3virgoenv}), with different and time-varying characteristics (amplitude and frequency contents). Detector characterization targets the smooth and usually stationary noise floor, which makes the envelope of the sensitivity curve. Beyond that, two main categories of noise artifacts are studied in detail as they can impact the performances of the instrument in detecting genuine \ac{gw} signals. The first one includes all noise transients, also called {\em glitches}, while the second one gathers all long-lasting noise excesses, the {\it spectral noises} (i.e. {\em lines} or {\em bumps} depending on their bandwidth type, narrow or wide, in the frequency domain).

To help understand the source of some of the noises that affect the Virgo sensitivity, hundreds of {\em auxiliary channels} monitor continuously the detector control systems as well as its local environment~\cite{EnvHuntVirgoO3}. They are also useful to vet the \ac{gw} candidates by assessing whether or not they seem to be of terrestrial origin.

\subsection{The LIGO-Virgo O3 run}

The third \ac{ligo}-Virgo Observing Run (O3) lasted about 11 months in total. It was divided into two parts: O3a, from April 1$^{st}$, 2019 to October 1$^{st}$, 2019; O3b, from November 1$^{st}$, 2019 to March 27$^{th}$, 2020, separated by a one-month commissioning break in October 2019. O3b should have lasted one more month but the worldwide Covid-19 pandemic forced the \ac{ligo} Scientific and Virgo Collaborations to end the data taking prematurely. A few days later, all three detectors were shutdown to cope with the various lock-down constraints.

The Virgo detector takes data in a configuration called {\em Science mode}. It corresponds to periods during which the instrument is controlled at its nominal working point, with that control stable and accurate enough to assume that the recorded data are of good quality and suitable for physics analysis. This assumption is checked in real time against online data quality checks and the corresponding dataset is further refined by offline studies to make the final Virgo dataset.

\subsection{Running the Virgo DetChar analysis tools}

All DetChar analyses rely on dedicated software frameworks, generically called {\em tools} in the following. Some have been designed and set up within Virgo to meet the goals of the DetChar group, while others have been developed partly or totally by \ac{ligo} colleagues. In fact, any DetChar tool can potentially be used by the three groups (Virgo, \ac{ligo}, and \ac{kagra}) thanks to the long-lasting collaborations among them.

More than 100 computing servers have been used in real-time during O3 to control and monitor the Virgo detector, run various data quality checks and perform specific DetChar tasks. Data are processed by the tools described in the following sections. Their outputs are included in the live data streams if they are available with a latency low-enough (about 15~s), or stored on disk otherwise. The end products of these analyses are converted into information for the control room and live summary plots that are updated with a latency of a few minutes at most, and regularly archived for reuse during offline analyses.

All this software framework is controlled using the \ac{vpm} software interface, that allows to configure, start/stop and monitor processes running on Virgo online servers. These include detector control, data transfer to and from Virgo, as well as the analysis of the reconstructed $h(t)$ stream by the online \ac{gw} search algorithms running in the \ac{ego} computing center. All actions performed using the \ac{vpm} interface are logged and recorded, in order to reconstruct as accurately as possible the running conditions at any given time, should this need arise.

The most important DetChar tools used by the Virgo group during the O3 run are classified in a few main categories depending on their usage or target: monitoring, generic data analysis, glitches and spectral noise investigations, or database management. Yet, they are not independent: they are often combined to characterize specific features of the detector, or to provide a complete overview of the quality of the Virgo data around a GPS time of interest. Such GPS ranges are called {\em segments}.
 
\begin{figure}[th!]
  \centering
  \includegraphics[trim=0 10 40 0, clip, width=\textwidth]{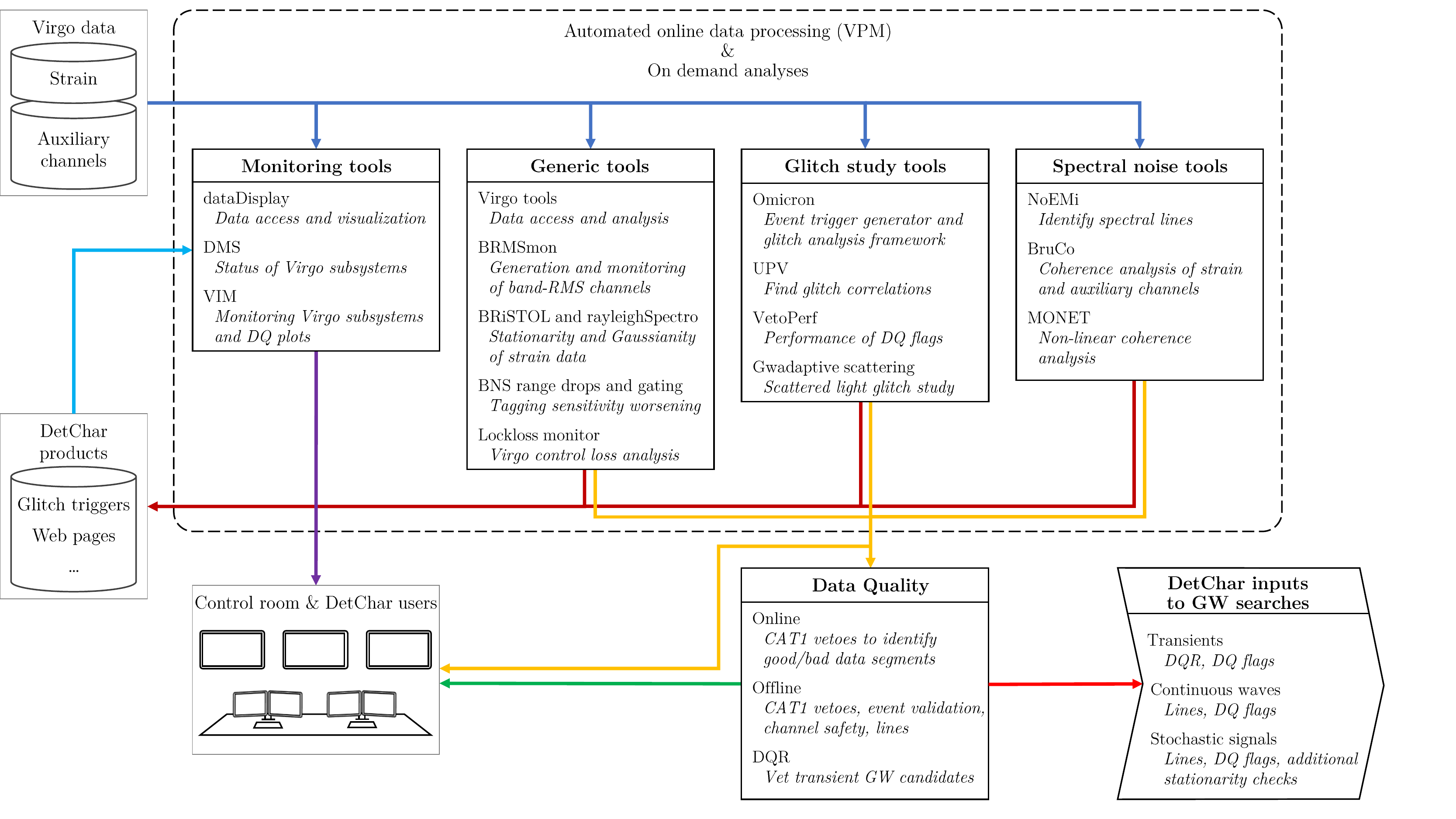}
  \caption{Flowchart of the various tools and monitors used for the Virgo DetChar studies during the O3 run --- the results of these analyses are presented in the companion article~\cite{O3DetChar_results}. The dataflow starts from the raw data acquired from the detector. These are then analysed by a wide set of DetChar tools, which in turn produce outputs of various types, used at different analysis levels. Either to monitor the detector and the data taking, or to construct data quality (in short ``DQ'') estimators, to be used by \ac{gw} search pipelines.}
  \label{fig:tools_flowchart}
\end{figure}

The flowchart in figure~\ref{fig:tools_flowchart} presents an overview of the main analyses carried out by the DetChar group and shows the corresponding tools described in the following.
The arrows follow the dataflow which starts from the detector raw data (top left corner) and goes all the way down to the final consumers of DetChar products: the on-duty crew in the Virgo control room, the broad community of DetChar users and the data analysts.

\subsection{Article contents}

This article categorizes the different DetChar tools used during the O3 run.
It aims at becoming the main reference for those which had not been documented yet,
while it puts all of them into perspective, by showing how they interact and
complement each other.

First, section~\ref{section:monitoring_tools} presents the main monitoring tools commonly used to make quick and basic analysis of the Virgo data, or to get a digest of the detector status and of its performance at any given time. Then, Section~\ref{section:generic_tools} highlights a wide set of DetChar-generic tools: a software layer to ease access to the Virgo data for the users; band-limited \ac{rms}-based estimates as quantities to identify noise contributions and follow their evolution over time; different methods to test the noise stationarity and Gaussianity, two hypothesis which are often explicitly or implicitly assumed by many data analyses; finally, some tools detecting and studying peculiarities during data taking phases, that is sudden drops of the sensitivity and losses of the detector global working point.

Section~\ref{sec:tools:glitch} focuses on transient noise bursts, or glitches. It includes well-established tools to identify them in time-frequency representations and to search for their origins. In addition, more recent scattered light monitors are described as well, as this kind of noise appears to be a real nuisance for all ground-based \ac{gw} detectors. The following section~\ref{section:spectral_noise_tools} is dedicated to the tools used to identify spectral noise features and to search for their origin. For the latter part of the investigations, coherence in the frequency domain is the key quantity used to connect spectral noise in the \ac{gw} strain $h(t)$ with auxiliary channels.

Then, the short section~\ref{section:common_tools} introduces two databases jointly used by \ac{ligo} and Virgo, one to gather data quality inputs for the analyses, and the other to store information about all \ac{gw} candidates found by those searches. Finally, section~\ref{subsection:DQR} presents the \ac{dqr}, the new DetChar framework developed specifically for the O3 run in order to vet the \ac{gw} candidates, in particular those found in low latency by the online data analyses.

Section~\ref{section:outlook} concludes this article by providing the outlines of the DetChar technical work to prepare the O4 run\footnote{The fourth \ac{ligo}-Virgo-\ac{kagra} Observing Run O4, is currently expected to start in Spring 2023.}, using all the experience accumulated during O3, and later on to analyse its dataset.

The main results of the Virgo DetChar analyses performed on data from O3 are presented in the companion article~\cite{O3DetChar_results}.

The abbreviations and acronyms used in this article are defined in a dedicated section at the end.


\section{Monitoring tools}
\label{section:monitoring_tools}

\subsection{dataDisplay}
\begin{figure}
  \center
  \includegraphics[width=0.8\textwidth]{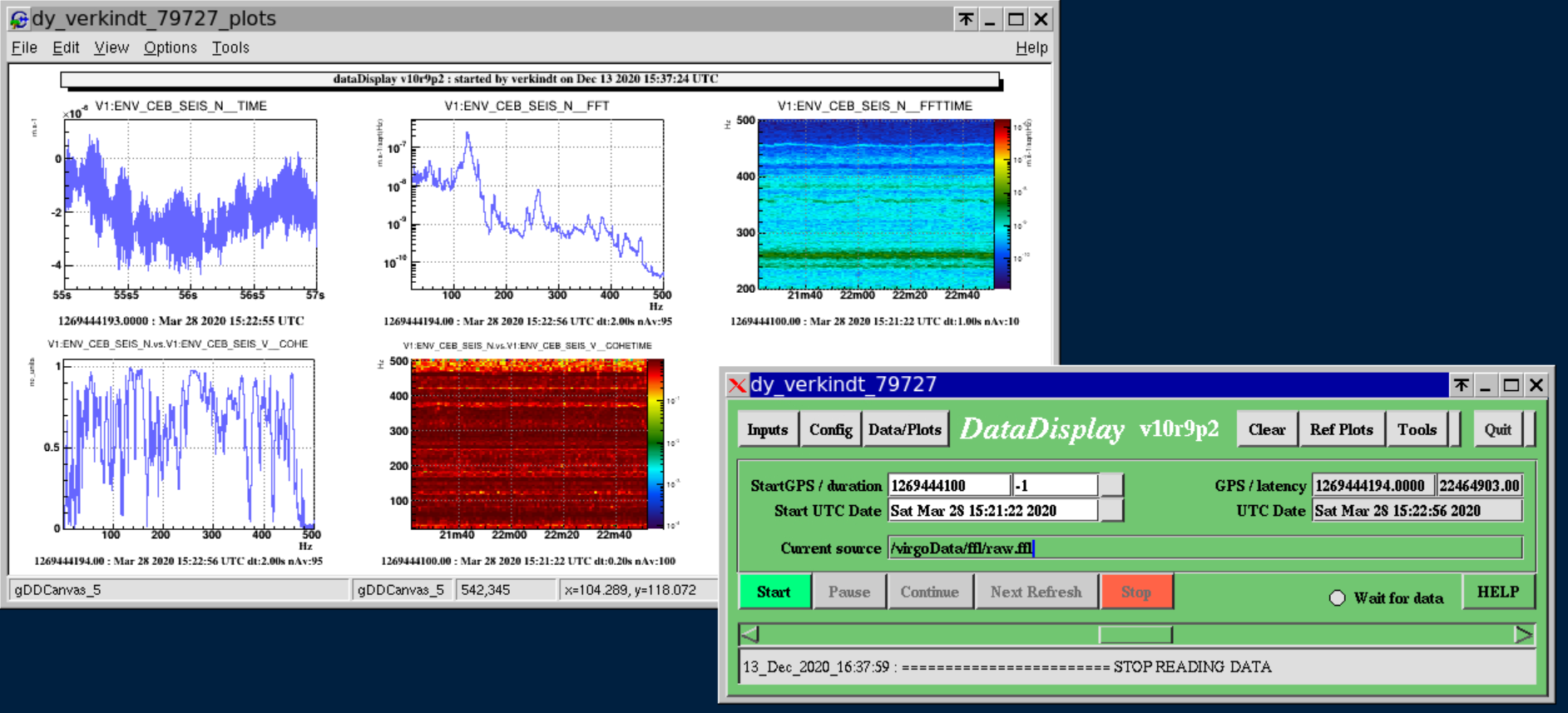}
  \caption{Example of the plots produced by the \texttt{dataDisplay} (left) and main panel of its graphical user interface (right).}
  \label{fig:dy}
\end{figure}

The \texttt{dataDisplay} software~\cite{datadisplay} allows the user to read (online or offline) Virgo data and to visualize various types of plots for all the channels available from the \ac{daq}.
For instance, it helps to investigate
quickly the time evolution of a noise artifact, the coherence between two control signals or the
time-frequency characteristics of a transient noise, etc.
It has been used extensively during the O2\footnote{O2 was the second \ac{ligo}-Virgo Observing Run. It started on November 30$^{th}$, 2016 only with the two \ac{ligo} detectors. Virgo joined O2 on August 1$^{st}$, 2017 and the three detectors took data jointly until August 25$^{th}$.} and O3 runs and all over the \ac{adv} detector commissioning in between.
Figure~\ref{fig:dy} shows an example of the \texttt{dataDisplay} interface and output.

\subsection{DMS: the Detector Monitoring System}\label{sec:tools:dms}
The \ac{dms}~\cite{dms1,dms2} provides a detailed live status of all the components that make the Virgo detector operate, from the hardware parts to the online software used to control the instrument, acquire the data and process it. It also includes the monitoring of environmental sensors from around the experimental areas. 
Each of the many \ac{dms} monitors uses a set of \ac{daq} channels, combines them by performing mathematical and logical operations on their outputs and produces a flag whose value can take four severity levels, each associated with a color for visual display. A web interface is used to display and browse the \ac{dms} monitor flags with a few second-latency, both in the Virgo control room and remotely.

\begin{figure}
  \centering
  \includegraphics[width=\textwidth]{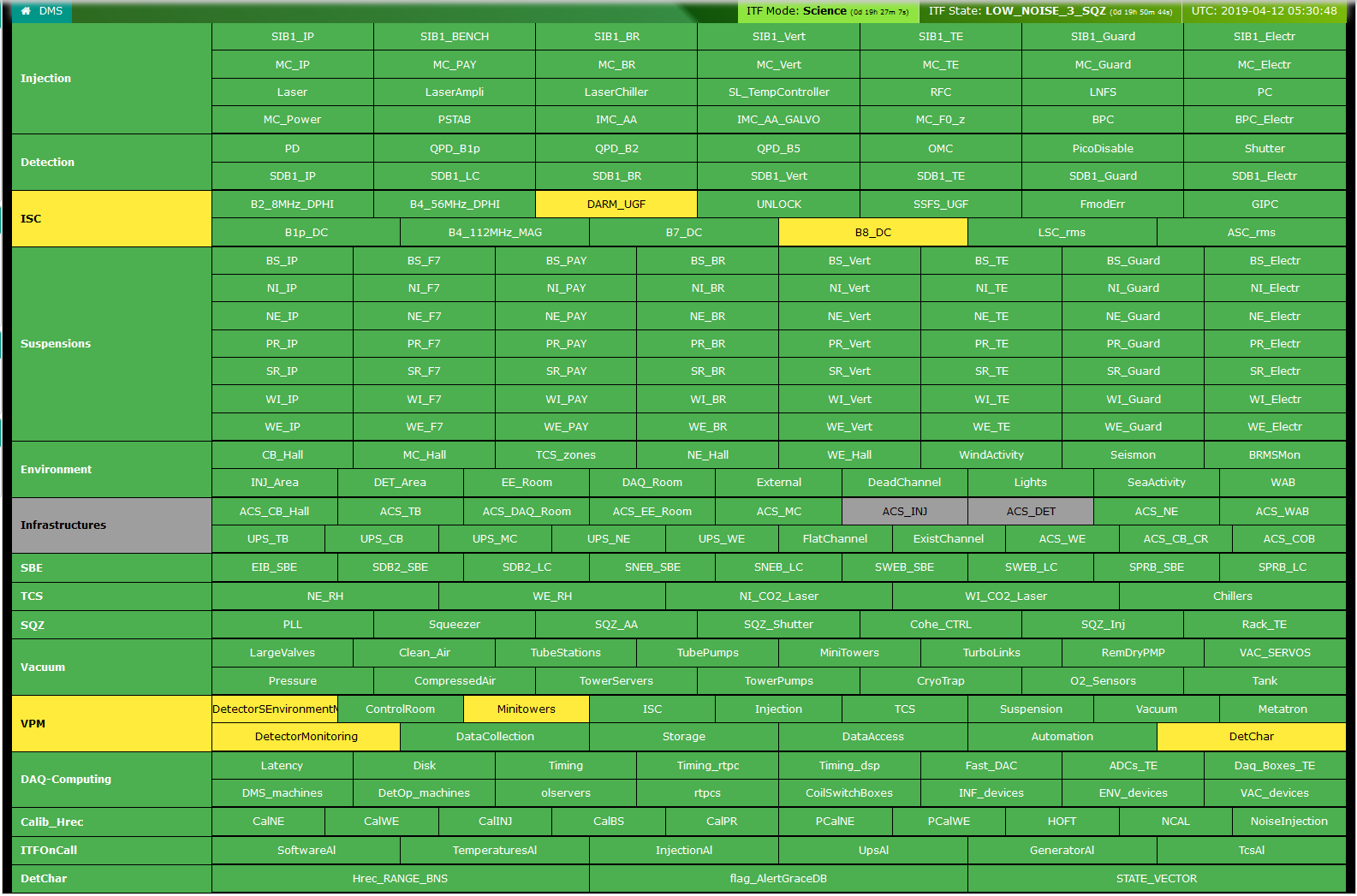}
  \caption{\ac{dms} snapshot closest in time to the GW190412 \ac{gw} event, showing the detailed status of the Virgo detector about 4~s after the arrival of that signal.
}
  \label{fig:DMS_playback_GW190412}  
\end{figure}

For instance, figure~\ref{fig:DMS_playback_GW190412} shows the Virgo detector status about 4~s after the detection of the \ac{gw} event GW190412~\cite{LIGOScientific:2020stg}. The \ac{dms} web interface looks like a checkerboard. Each row, labeled in the most-left column, corresponds to a different part of the instrument (mirror suspensions, vacuum system, etc.). Moving to the right of the screenshot, that part is broken down in smaller sets that are each associated with a cell on the web interface. Each cell can contain many \ac{dms} flags and its color reflects the highest severity among all these flags (green $\leftrightarrow$ no alarm; yellow $\leftrightarrow$ warning; red $\leftrightarrow$ alarm (not present on that particular snapshot); grey $\leftrightarrow$ some information is missing). Clicking on a cell gives access to the flag individual information: their values and associated severities.

In addition to the live global detector status, a new \ac{dms} archival system has been set up for the O3 run: complete \ac{dms} snapshots are taken every ${\sim} 10$~s and archived. They can be retrieved later at any time, by running a playback application that uses the same interface as the live \ac{dms}.
This functionality is particularly convenient to check the status of the detector a posteriori, when a \ac{gw} candidate or a particular feature in the data have been identified.
 
\subsection{VIM: the Virgo Interferometer Monitor}\label{sec:tools:VIM}
\begin{figure}
  \center
  \includegraphics[width=0.95\textwidth]{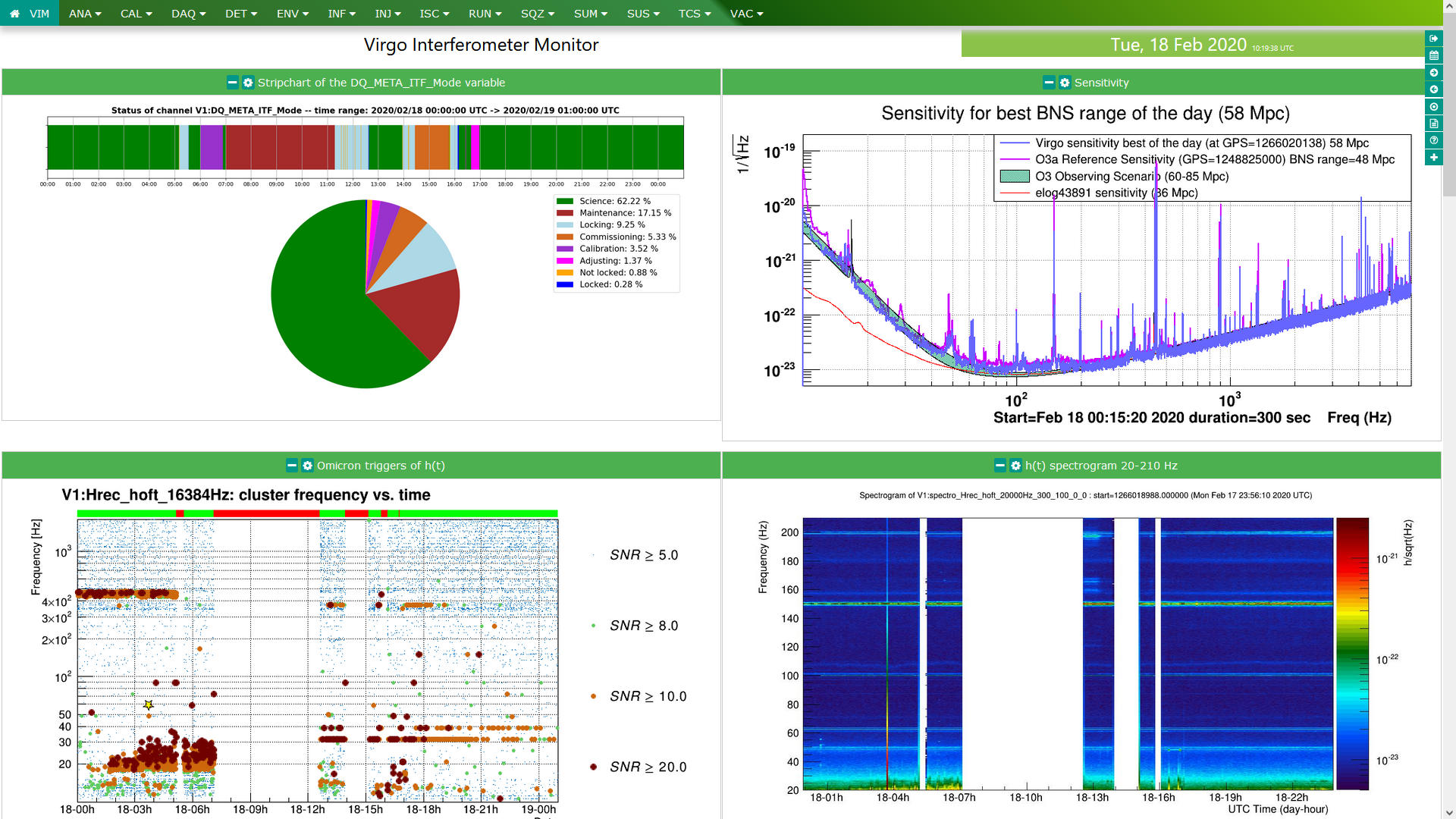}
  \caption{Screenshot of a \ac{vim} web page displaying information about the Virgo detector on Tuesday February 18$^{th}$, 2020. Top left plot, stripchart of the detector status: the weekly maintenance, preceded by a planned calibration and followed by a short commissioning period, interrupts the data taking that restarts in the evening. Top right plot: daily sensitivity compared with references. Bottom left plot: glitch monitoring provided by the \texttt{Omicron} analysis described in section~\ref{sec:tools:glitch:omicron}. Bottom right plot: spectrogram of the \ac{gw} strain $h(t)$ in the 20-210~Hz frequency range.}
  \label{fig:o3_vim}  
\end{figure}

The \ac{vim}~\cite{hemming2016,verkindt2019} manages a collection of automated scripts that update every few minutes a wide set of plots and tables; all these monitoring products are permanently archived on a daily basis. A web interface allows users to browse that database, both for live monitoring of the experiment and for offline investigations.
\ac{vim} is an essential tool that provides a direct access to a detailed status of the various Virgo detector components and of related frameworks, such as calibration and online data processing, data transfer or online data analyses.
A snapshot of the \ac{vim} web interface is shown in figure~\ref{fig:o3_vim}.

\section{Generic tools}
\label{section:generic_tools}
\markboth{\thesection. \Sectionname}{}

\subsection{The VirgoTools utilities}
In-depth studies of a particular feature observed in the data or analyses scanning a significant fraction of the dataset require the use of dedicated software. Common and key building blocks of these codes have access to the \ac{daq} channels and to the detector component configurations. Thus, dedicated packages have been developed over the years to provide simplified and generic interfaces to these data: they rely on low-level core packages like the \texttt{FrameLib} software library~\cite{FrameLib} but calls to these functions are hidden to the users. These packages interact with the software, hardware and data of the Virgo interferometer: they are widely used within the collaboration, from daily use in the control room to DetChar studies. The two main collections of such functions are \texttt{PythonVirgoTools}~\cite{PythonVirgoTools} and \texttt{MatlabVirgoTools}, targeting Python and Matlab developers, respectively.

\subsection{Computing Band-limited RMS}\label{sec:tools:monitoring:brmsmon}
\acp{brms} of \ac{daq} channels computed in specific frequency ranges after Fourier-transforming the time series are useful indicators for transient disturbances or new features in the data~-- see Eq.~\ref{eq:defBRMS} in Sec.~\ref{appendix:BRiSTOL} below. For instance, low-frequency \ac{brms} of seismometer data allow to separate different contributions to the seismic noise at \ac{ego}~\cite{o3virgoenv}. Going from low to high frequencies, one can isolate successively: distant and potentially strong earthquakes; sea activity on the Tuscany coastline; anthropogenic contributions with day/night and weekly periodicities; finally, on-site activities. In addition, \ac{brms} is used to monitor the excitation of the violin modes, i.e. the resonances of the mirror suspensions.

In Virgo, various software frameworks can compute \ac{brms}. One worth-mentioning is \texttt{BRMSMon}, a dedicated software that is widely used by the environmental monitoring team and in data quality studies. In addition to generating \ac{brms}, \texttt{BRMSMon} can compare their values to thresholds (either fixed or adaptive) and logically combine the outputs of these comparisons into binary channels called {\em flags}. For instance, assuming a collection of 9 sensors installed in
different \ac{ego} buildings, one can create a flag that is active (value equal to 1) if at least 5 of these 9 sensors exceed their own threshold and inactive (value 0) otherwise. The \texttt{BRMSMon} output channels, sampled at 1~Hz, are included in the \ac{daq}.

\label{subsubsection:BLRMS}

\subsection{Testing stationarity and Gaussianity}\label{sec:tools:Bristol}
Several tools have been implemented to perform statistical tests to verify the stationarity and Gaussianity of the data.
These properties are indeed the typical assumptions at the base of most of the statistical analyses of \ac{gw} search pipelines, such as \texttt{MBTA}~\cite{Aubin:2020goo}, \texttt{PyCBC}~\cite{PyCBCLiveO3} and \texttt{GstLAL}~\cite{messick2017analysis} that are based on the matched filter technique~\cite{Davis1989,Abbott_2020}.
Moreover, the onset of a non-stationary behavior of the detector can be the symptom of some hardware malfunction
or some contamination from environmental noises. In any case, it requires prompt investigations of its causes and, possibly, the actuation of adequate mitigation strategies.

\subsubsection{BRiSTOL}\label{appendix:BRiSTOL}

The detector output records can be described as the realization of a stochastic process.
This process is said to be (strict-sense) stationary if its statistical properties are left unchanged by shifts in time, which, in most practical situations, allows us to estimate them from a sufficiently long realization of the process (\emph{ergodic hypothesis}), and this estimate does not depend on when, in time, it has been performed.
For most of the analysis, it is sufficient to consider a \emph{weaker} form of stationarity that involves only the first distribution moments, namely the mean and the covariance function.
Moreover, if the process is stationary, we can define the \ac{psd} of the process as the Fourier transform of its covariance~\cite{Abbott_2020}.

The \ac{bristol} statistical test~\cite{DiRenzo:2020} verifies the hypothesis of weak-sense stationarity by verifying that subsequent \ac{psd} estimates, in predefined frequency bands, are compatible with the same probability distribution. 
The corresponding test statistics are based on a set of \ac{brms} time series, estimated on an equal number of bands:

\begin{equation}\label{eq:defBRMS}
\mathit{BRMS}_t(b) = \sqrt{\int_{f\in b} \hat{S}_t(f)\,df}, \qquad \mathrm{for} \quad b\in\left\{[f_1 ^{\max}, f_1 ^{\min}],\ldots,[f_K ^{\max}, f_K ^{\min}] \right\}
\end{equation}

for data $x_{t_n}$ recorded at Nyquist rate $f_S$, $t_n = t+n/f_S$, where $\hat{S}_t(f)$ is a \ac{psd} estimate referred to time $t$, and obtained with the \emph{periodogram} method~\cite{schuster1898}:

\begin{equation}\label{eq:periodogram}
    \hat{S}_t(f_k)=\frac{1}{N}\bigg|\sum_{n=0} ^{N-1} x_{t_n} e^{-2\pi i\, nf_k/f_S}\bigg|^2,\qquad f_k=\frac{k\,f_S}{N}, \quad\text{for}\quad k=0,\ldots,N-1
\end{equation}

Two modifications have been implemented to make equation~\eqref{eq:defBRMS} more suitable for the study of transient noise, in particular to identify ``slow non-stationarities'', namely changes in the statistical properties of the data over time scales longer than a second. 
First, frequencies corresponding to spectral lines (refer to section~\ref{sec:noemi} for more details) have been removed from the integral in~\eqref{eq:defBRMS}.
These lines are narrow features in the \ac{psd} of the data, originating from resonances in various parts of the interferometer and their harmonics.
Their intensities can be orders of magnitude larger than the neighboring noise floor.
Hence, if a line is present in a band where we are about to compute the \ac{brms}, it is likely to dominate the final estimate, and also the corresponding fluctuations, preventing us from probing the features of the underlying noise floor.
To remove these lines, we identify them with an algorithm similar to the one developed for the \ac{noemi} pipeline~\cite{Accadia_2012}, and based on the \emph{prominence} of their \ac{psd}~\cite{acernese2005simple}.

Second, also glitches are typically removed from the \ac{brms} time series.
As these glitches can manifest at a rate of about 10 per minute
(see O3 glitch rates in~\cite{O3DetChar_results}), every data segment longer than a few seconds is likely to contain one of them, hence leading us to reject stationarity over such time scales.
Dedicated algorithms, based on excess noise identification, are typically used to find them, and will be described in section~\ref{sec:tools:glitch}.
However, these algorithms are in general not sensitive to slower non-stationarities.
To make \ac{bristol} specifically targeted at the latter, we have then proceeded to identify glitches on the \ac{brms} time series by means of an algorithm based on a rolling \emph{median absolute deviation} (defined as the median absolute difference from the median), and to remove the corresponding data points from the analysis.

For each frequency band, the resulting modified \ac{brms} time series are divided into ``chunks'' of equal duration, which are used to test stationarity.
This hypothesis is tested by means of a two-sample Kolmogorov--Smirnov test~\cite{kolmogoroff1941confidence} for each pair of consecutive chunks, and the corresponding $p$-values are compared to a test significance $\alpha$ (to be decided in advance).
The (\emph{null}) hypothesis of stationarity is rejected when a $p$-value is less than or equal to $\alpha$, meaning that the estimates in the two chunks are drawn from different distributions and the underlying process has changed.

There are two advantages in using \ac{brms}-based quantities.
First, averaging over the frequencies of each band has a similar variance reduction effect than the means in Welch's \ac{psd} estimation method~\cite{welch1967}.
This in turn allows a finer time resolution while maintaining a moderate variance for the test statistics, that is, the empirical distribution of the \ac{brms}.
Second, the various non-stationarities typically manifest in specific frequency bands, closely related to the noise source that generated them.
For example, the main harmonic of scattered light is usually visible below $30~\mathrm{Hz}$; non-linear and non-stationary couplings of the angular controls with the $150~\mathrm{Hz}$ harmonic line are characteristic of a tight region around it, etc.
So, without losing much of resolution, we can perform the noise characterization directly on these bands instead of on each frequency bin comprising the spectrum of the signal.

\begin{figure}[!t]
	\centering
	\includegraphics[width=.85\textwidth]{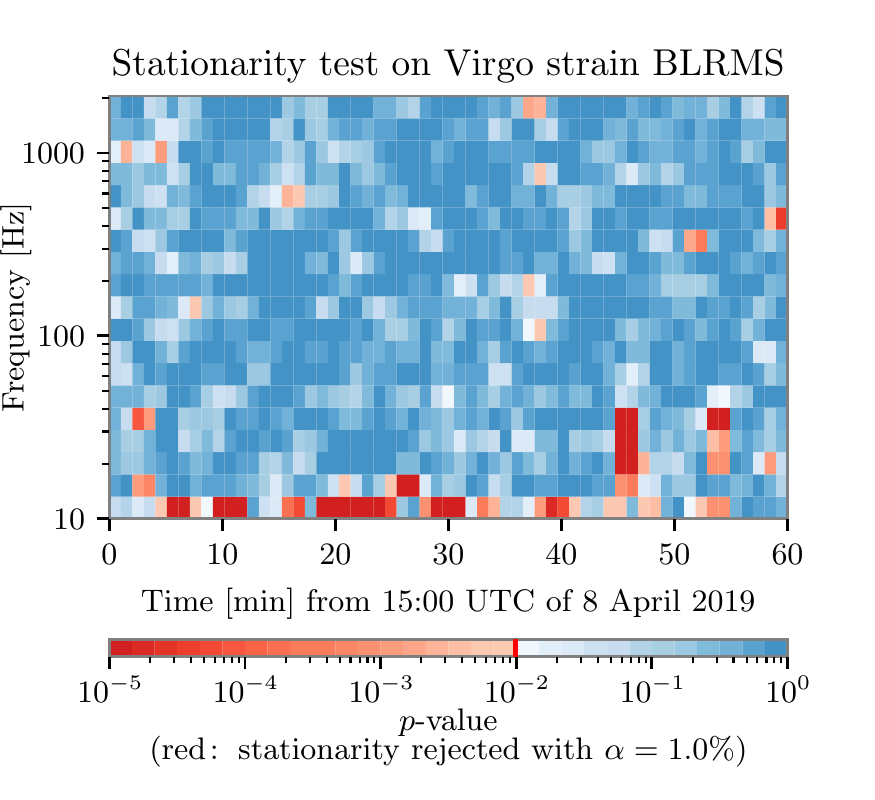}
	\caption{Example of stationarity time--frequency map obtained with \ac{bristol}, where a significance $\alpha=1\%$ has been chosen and regions rejecting the stationarity hypothesis are colored in shades of red.
		\label{fig:bristol}}
\end{figure}

The output of \ac{bristol} can be visualized as a time--frequency map where, for each bin, the value of the test statistic $p$-value is reported.
An example of such a map is shown in figure~\ref{fig:bristol}, where different color palettes are used to highlight those regions where stationarity should be rejected (red) and where not (blue).
The time resolution of the test is given by the duration of each chunk and that of the \ac{psd} estimates, typically one minute and one second respectively; that in frequency is determined by the band division of the spectrum for computing the \ac{brms}, which is conveniently done choosing exponentially spaced frequency intervals.

This tool has been developed in the commissioning phase preceding O3, and has been used during the run as well, to assess the quality
of the data as part of the event validation procedure (refer to~\cite{O3DetChar_results} for more details).

\subsubsection{rayleighSpectro - Gaussianity test}\label{appendix:Rayleigh}

Similarly to what was discussed for the stationarity hypothesis, Gaussianity has to be tested separately in the different regions of the spectrum where noise sources can show up. 
\texttt{rayleighSpectro}~\cite{verkindt_spectro}, based on the \emph{Rayleigh test}~\cite{finn2001}, does this by means of a consistency check on the \ac{psd} estimated from the data with what is expected for stationary Gaussian noise.
Indeed, if the data is compatible with the hypothesis of Gaussianity, the periodogram estimator in equation~\eqref{eq:periodogram}
is asymptotically (with $N$ large) described by an \emph{exponential distribution} of parameter $S(f_k)^{-1}$~\cite{KOKOSZKA200049}, where $S(f_k)$ is the processed \ac{psd}.
The corresponding \ac{asd} estimator, obtained as the square root of equation~\eqref{eq:periodogram}, is described by a Rayleigh distribution with parameter $\sqrt{S(f_k)/2}$.
The scaling property of this distribution can be used to construct consistency tests.
For example, the standard deviation of the \ac{asd} estimates obtained on non-overlapping segments provides an estimator of the standard deviation of this variable, which equals to $\sqrt{(4-\pi)\,S(f_k)/2}$.
Similarly, the mean of these estimates provides an estimator of the mean: $\sqrt{\pi S(f_k)/2}$.
The ratio of these two quantities gives a test statistic that, at each frequency $f_k$, is \emph{asymptotically} equal to
\begin{equation}\label{eq:Rayleigh_test_statistic}
	\frac{\sqrt{4-\pi}}{\sqrt{\pi}}\simeq 0.52.
\end{equation}
The actual value of the previous quantity for a finite number of averages and the corresponding critical values
for performing statistical tests have been computed in~\cite{verkindt_spectro, direnzo2020ragout}.

If the noise is not Gaussian, or its properties has changed while estimating the standard deviation and mean of its \ac{asd}, the resulting test statistic will take values different than what reported in~\eqref{eq:Rayleigh_test_statistic}.
This means that the Rayleigh test is sensitive to both non-Gaussianities and non-stationarities in the data.
Smaller values of the statistic are associated with data having smaller fluctuations than those expected for a Gaussian process; {spectral lines} usually behave in this way.
Larger values are instead typical of non-stationary noises, such as glitches, that produce a larger variance of the \ac{asd} estimates.

By dividing the data into chunks of duration $\Delta t$, one can obtain a time--frequency map,
similar to a spectrogram, showing with time resolution $\Delta t$ the frequencies and times where the data significantly depart
from the expected value of equation~\eqref{eq:Rayleigh_test_statistic}.

The output of \texttt{rayleighSpectro} are included in \ac{vim} and also used in the \ac{dqr} for event validation (see~\cite{O3DetChar_results} for more details).

The interplay between this tool and \ac{bristol} for the assessment of the stationarity and Gaussianity of the data is the following.
\ac{bristol} assesses where the data is compatible with the hypothesis of wide sense stationarity, that is, the second order moments (i.e. the covariance or the \ac{rms}) are left unchanged by shifts in time.
This corresponds also to a \emph{strong sense stationarity} (the invariance of the probability distribution of the noise) if the data is also Gaussian, that is, completely characterized by its mean and covariance functions, as tested by \texttt{rayleighSpectro}.
Conversely, deviations from these assumptions can be tested independently.
This is useful for example with regions corresponding to spectral lines that are typically stationary in very good approximation but not Gaussian.

\begin{figure}[!t]
	\centering
	\includegraphics[width=.85\textwidth]{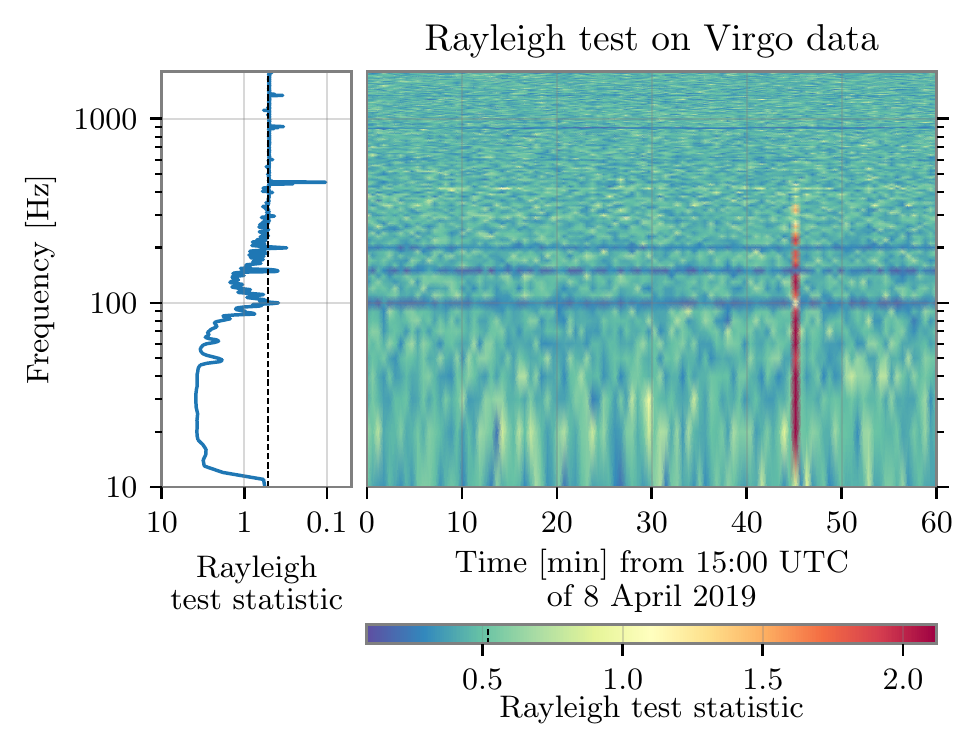}
	\caption{Example of application of the Rayleigh test, where the blue line in the left panel is the test statistic estimated over the entire hour of data, while the color map corresponds to \ac{asd} estimates over 10~s of data. The vertical dashed black line on the left plot indicates the 0.52 limit value. 
	See text for details.
		\label{fig:rayleigh}}
\end{figure}

Figures~\ref{fig:bristol} and~\ref{fig:rayleigh} show examples of the application of these two tools to 1~h of data at the beginning of O3a. In the first plot, \ac{bristol} highlights many slow non-stationarities at frequencies up to about 20~Hz, most likely due to high microseismic activity, as well as a loud glitch at about 15:45~UTC. This one is clearly identified by the Rayleigh test with values of the test statistic larger than what is expected for stationary and Gaussian noise. Moreover, in the color map of figure~\ref{fig:rayleigh}, spectral lines, in particular those 
associated with the 100, 150 and 200~Hz harmonics of the mains (the European power grid frequency is 50~Hz), 
are highlighted in blue, corresponding to values of the test statistic smaller than the asymptotical limit in the Gaussian case, given in equation~\eqref{eq:Rayleigh_test_statistic}. 
In the left-hand side panel of the same image, the 450~Hz frequency of the main test masses violin modes, and its first harmonic at about 900~Hz, are highlighted as well.

\begin{figure}[!htb]
	\centering
	\includegraphics[width=.9\textwidth]{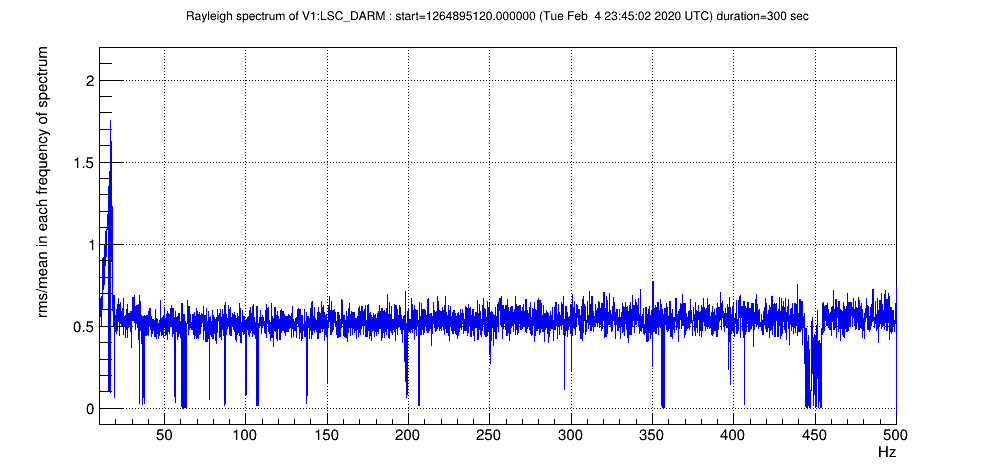}
	\caption{Example of Rayleigh spectrum averaged over 300~s, where any bin above 0.52 is a potential non-stationary or non-Gaussian noise present during those 300~s. Values well below 0.52 correspond to persistent frequency lines. 
	}
	\label{fig:rayleigh_dv}
\end{figure}

Figure~\ref{fig:rayleigh_dv} shows another example of a Rayleigh spectrum from O3, corresponding to a period during which a transient noise was present for several minutes between 10~Hz and 20~Hz.

\subsection{Monitoring BNS range drops and gating data}
The \ac{bns} range is the average distance up to which the merger of a \ac{bns} system can be detected. The average is taken over the source location in the sky and the \ac{bns} system orientation, while a detection is defined by convention as a \ac{snr} of 8 or above. This quantity is related to the live detector sensitivity and can be used as a summary statistic to monitor the overall performance of the instrument in relation to transient searches.

Two useful high-level data quality monitors are based on the \ac{bns} range\footnote downwards excursions. One tags \ac{bns} range {\em drops}, that are significant and sudden decreases in this value, usually flagging data quality problems. The other automatically generates (logical) {\em gates} that are applied on the \ac{gw} strain channel to smooth out to zero the data that are affected by a strong noise transient. \ac{bns} range drops and gates are related but not equivalent depending on the frequency contents of the noise burst.

\subsubsection{\ac{bns} range drops}

A \ac{bns} range drop means that the live sensitivity of the detector is degrading significantly, at least in a given frequency band, possibly in the entire bandwidth of the instrument. 
Therefore, it is important to identify transient sensitivity worsenings and investigate their causes.
\ac{bns} range drops are very diverse: the decrease goes from a few percents to almost the full range, while the drops can last from a few seconds to minutes.
 
\begin{figure}
  \center
  \includegraphics[width=0.8\textwidth]{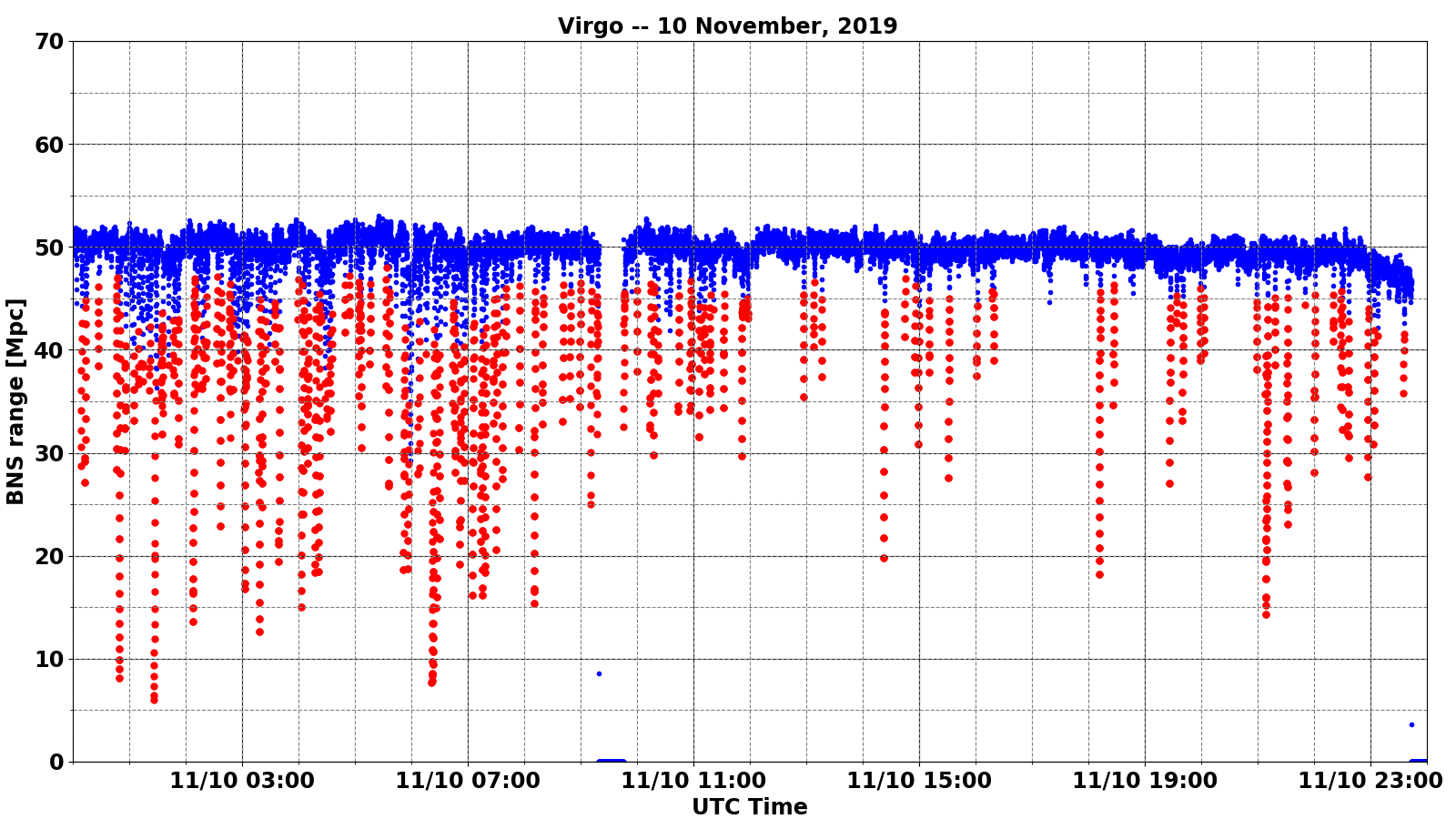}
  \includegraphics[width=0.8\textwidth]{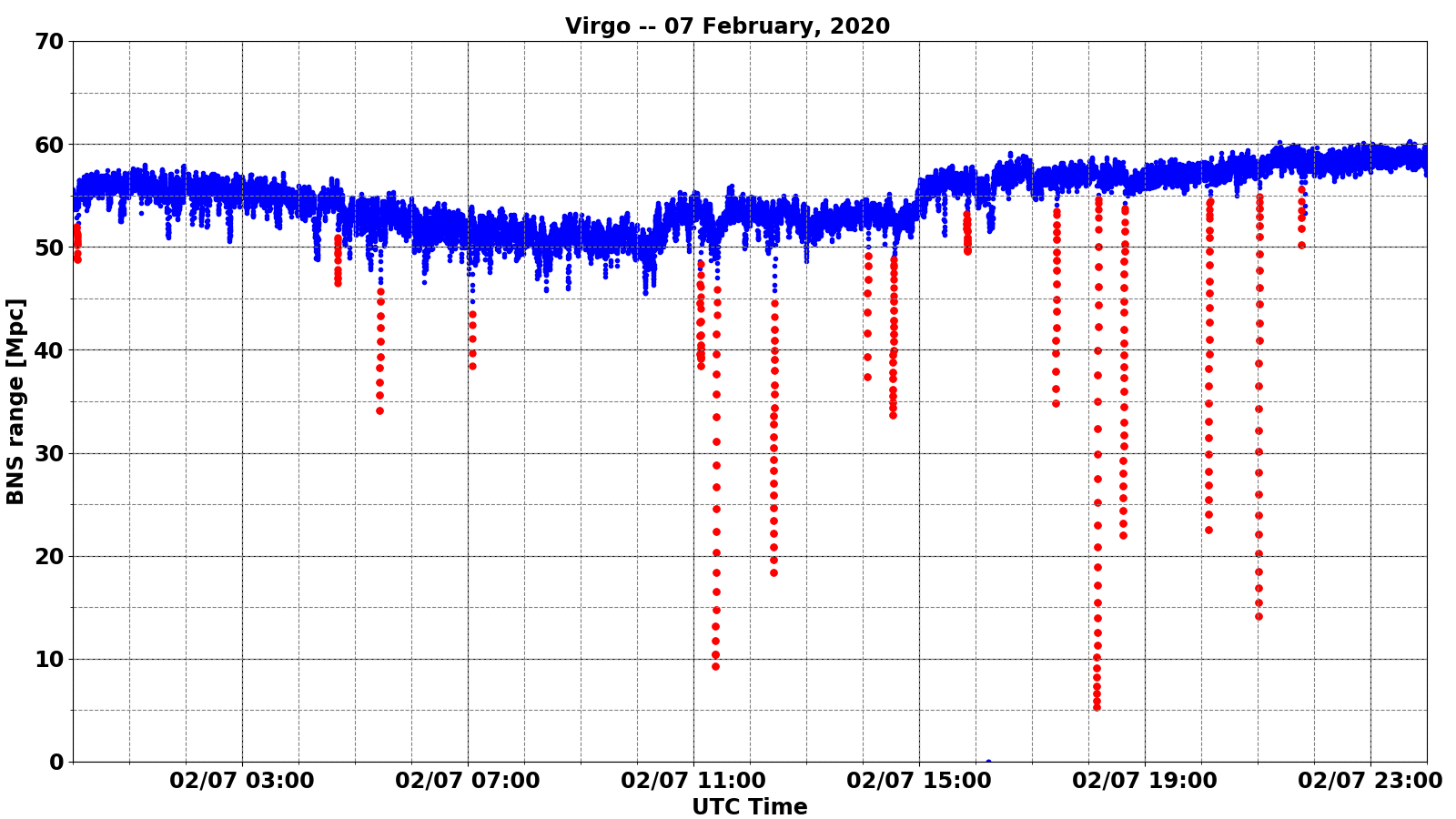}
  \caption{Performance of the \ac{bns} range drop locator during two days of O3. Top plot: November 10$^{th}$, 2019, a day during which the duty cycle was quite high but the data taking conditions were not stable; many glitches and consequently \ac{bns} range drops were observed, mostly due to the laser power stabilization system in the morning and to a worsening of the weather conditions starting from the afternoon. Bottom plot: February 7$^{th}$, 2020, a day with no global control loss but a \ac{bns} range baseline varying over time; actions took place during the afternoon to improve the Virgo performance, leading to visible improvements of the \ac{bns} range in steps.
The blue traces show the range vs. time, while the red dots show the drops that have been identified.
In both cases the \ac{bns} range drop locator is able to identify most, if not all, significant drops.
}
  \label{fig:BNS_range_drops}
\end{figure}

During O3, \ac{bns} range drops were detected using an absolute threshold on the live value of that quantity. After the end of the run, adaptive methods able to follow the natural evolution of the \ac{bns} range and to locate all significant drops have been developed.
Figure~\ref{fig:BNS_range_drops} shows examples of the output of the adaptive \ac{bns} range drop locator running on O3 data.

\subsubsection{Gates}

If not removed from data, noise bursts can pollute the estimation of the noise spectrum for several seconds, hence limiting the sensitivity of the \ac{gw} search algorithms during that period. In Virgo, this problem is mitigated online by gating out (i.e. smoothing out to zero) glitchy chunks of data. The gating algorithm triggers on significant \ac{bns} range drops: at least 40\% below its median value, computed over the last 10~s.
On both sides of the gate, a weight is applied on the $h(t)$ strain channel during 10/32$^{th}$ of a second, varying smoothly from 1 to 0 (0 to 1) before (after) the gate. The online gated $h(t)$ strain channel is included in the \ac{daq} alongside the ungated one, and \ac{gw} searches are free to use one or the other stream as input.

As gating is based on $h(t)$ variations, gated data cannot simply be removed from the physics-analysis  dataset as this procedure could flag real \ac{gw} signals, for instance loud high-mass binary black hole mergers. On the other hand, gating information can be used in a statistical way to help identify potential periods of bad data quality characterized by frequent gating usage. This can be measured using both the density of gates (number of gates per unit of time) and the fraction of the wall-clock time that is gated out.

During O3, the gating algorithm has produced more than 13,000 gates (corresponding to a few tens per day in average), adding up to about 4~h of gated data in total. The gate mean duration was around 1.1~s while the median was around 0.8~s, meaning that most gated glitches were very short as 20/32$^{th}$ of a second were always added to the measured glitch duration to transition from non-gated data to the gate itself and back. The longest gate was about 10~s.

The segments that have been vetoed for offline data analyses~\cite{O3DetChar_results} excluded from this online Science dataset led to a removal of about 20\% of the gates and of about 30\% of their total duration. However, this procedure only removed about 0.2\% of the Virgo O3 dataset. As expected, more gates were generated when the quality of the data was degraded. Going one step further by requesting in addition that the baseline \ac{bns} range be greater than 35~Mpc. This excludes more than 50\% of the remaining gates and more than 60\% of the gated times, while that cut would remove about 1\% of the data from the final dataset. Gates are generated more often when the data taking conditions are sub-optimal.

Finally, one can associate all gates with a glitch detected by \texttt{Omicron} (see section~\ref{sec:tools:glitch:omicron}) whereas the opposite is not true: there are many glitches that have no impact on the \ac{bns} range. These glitches have a frequency range that is outside of the Virgo bandwidth for \ac{bns} \ac{gw} waveforms: either because there is no significant signal contribution expected in that frequency range, or because the noise level is high enough to make that range contribute little if anything at all to the \ac{bns} range.

\subsection{Monitoring global Control losses}
Losses of the global working point of the Virgo interferometer (the mandatory configuration sensitive to passing \acp{gw}) do not just interrupt the data taking: they decrease the overall duty cycle as few tens of minutes are needed after such events to restore the conditions for taking good-quality data sensitive to the passing of \acp{gw}.
Therefore, categorizing control losses is important to understand their main causes and to get alerted when a new class of control losses appears, or when an already known category becomes more frequent.

An extensive offline study of the global control losses in Science data-taking mode during the O3 run has lead to the identification of the root cause of the control losses in most cases~\cite{o3virgoenv}. The experience gained with 
this work will be useful for the pre-O4 commissioning phase (noise hunting) and the subsequent data taking periods in two ways. First, the categories identified during O3 will be reused as a starting point to investigate new control losses. Then, an online monitor will analyse these global control losses within minutes of their occurrence; it will provide a set of automated plots for further human diagnosis and possibly point out their probable cause. This framework is currently under development and will reuse the approach (if not the proper software infrastructure) of the \ac{dqr} (see section~\ref{subsection:DQR}).


\section{Glitch identification and characterization tools}\label{sec:tools:glitch}
\markboth{\thesection. \Sectionname}{}

\subsection{Omicron}\label{sec:tools:glitch:omicron}
The \texttt{Omicron}~\cite{omicron-softx} search algorithm is used to detect and characterize transient noises. The data is processed using a $Q$ transform~\cite{Brown:1991} which consists in decomposing a time series $x(t)$ onto a generic basis of complex-valued sinusoidal Gaussian functions centered on time $\tau$ and frequency $f$:

\begin{equation}
  X(\tau, f, \sigma_t) = \int_{-\infty}^{+\infty}{ x(t) \frac{W}{\sigma_t\sqrt{2\pi}}\exp{\left[-\frac{(t-\tau)^2}{2\sigma_t^2}\right]} e^{-2i \pi f t}dt}.
  \label{eq:qtransform}
\end{equation}

\begin{figure}[htb!]
  \center
  \includegraphics[width=0.9\textwidth]{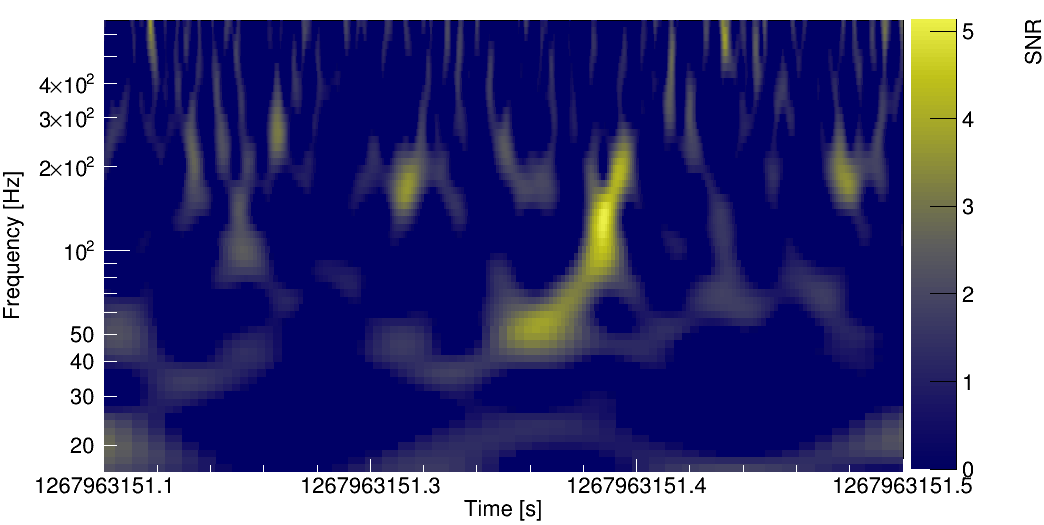}
  \caption{Spectrogram of GW200311\_115853~\cite{GWTC3} in the Virgo detector, as measured by \texttt{Omicron}. The time-frequency plane is tiled fixing $Q=6.4$.}
  \label{fig:gw200311}  
\end{figure}

This transformation is a modification of the standard short Fourier transform in which the analysis window size $\sigma_t$ varies inversely with the frequency and is characterized by a quality factor $Q$: $\sigma_t=Q/(\sqrt{8}\pi f)$. The parameter space $(\tau, f, Q)$ is tiled to guarantee both a high detection efficiency and an optimized processing speed. The noise of the input signal $x(t)$ is whitened prior to the $Q$ transform such that all noise frequencies have the same weight. This is done through the normalization factor $W$ which includes an estimate of the local stationary noise, such that the $Q$ transform coefficient $X$ directly measures the \ac{snr} associated to each individual tile $(\tau, f, Q)$. A glitch in the data is detected by \texttt{Omicron} as a collection of tiles with high-\ac{snr} values. An \texttt{Omicron} glitch is characterized by a set of parameters $(\tau, f, Q)$ given by the tile with the highest \ac{snr} value. \texttt{Omicron} offers a two-dimensional representation of glitches where the \ac{snr} distribution of tiles is plotted in one or several $Q$ planes.
\ac{gw} events can also be visualized with \texttt{Omicron}: see figure~\ref{fig:gw200311}.

\subsection{Use-Percentage Veto}\label{sec:tools:glitch:upv}
The \ac{upv} algorithm~\cite{Isogai:2010zz} has been developed to detect and characterize noise correlations between two glitch data samples; one derived from the \ac{gw} strain channel $h(t)$ and the other derived from an auxiliary channel. The algorithm tunes, considering \texttt{Omicron} triggers of a given auxiliary channel, a signal-to-noise ratio threshold such that, when a trigger is above threshold, there is a high probability to find a coincident glitch in the $h(t)$ data. In O3, the Virgo data were processed with the \ac{upv} algorithm on a daily basis to support the noise characterization effort; some auxiliary channels were identified by \ac{upv} as exhibiting glitches correlated with $h(t)$ glitches, providing hints about the noise coupling in the detector.

\subsection{VetoPerf}\label{sec:tools:glitch:vetoperf}
The \texttt{VetoPerf} analysis tool measures the performance of a data quality flag. A data quality flag is defined as a list of time segments targeting transient noise events. \texttt{VetoPerf} counts the number of the $h(t)$ triggers detected by \texttt{Omicron} which are coincident with the data quality flag time segments. From this, it derives performance numbers and produces diagnostic plots characterizing that data quality flag.

\subsection{Scattered light monitor}
Scattered light is a non-linear, non-stationary noise affecting the sensitivity of the interferometer in the \ac{gw} detection frequency band. As adaptive algorithms such as \ac{emd}~\cite{huang1998empirical,huang1999new,yang2009analysis} are suitable for the analysis of non-linear, non-stationary data, they can be used to quickly identify optical components which are sources, i.e. culprits, of scattered light~\cite{Valdes_2017}. As part of the detector characterization effort, a tool was developed and applied to Virgo O3 data with the aim of identifying culprits of scattered light in the \ac{darm} \ac{dof} of the detector~\cite{Longo_2020sc}. The tool employs the recently developed time varying filter \ac{emd} algorithm (tvf-\ac{emd})~\cite{Li_2017} which gives more accurate results compared to \ac{emd}~\cite{Longo_2020sc}. When scattered light is affecting the detector, arches show up in \ac{darm} spectrograms. The arches frequency and their time of occurrence is given by the so called predictor (measured in Hz):

\begin{equation}
f_{arch}(t)=2\frac{|v(t)|}{\lambda},
\label{predictor}
\end{equation}

where $v(t)$ is the velocity at which the optical component is moving and $\lambda$ is the laser wavelength. Equation~\eqref{predictor} is computed using the position data of several optics of the detector, such as for example the \ac{sweb}. Having obtained predictors for several optical components, the tool computes the instantaneous amplitudes $IA(t)$, i.e. the envelope of \ac{darm} oscillatory modes which are extracted by tvf-\ac{emd}. $IA(t)$ can then be correlated with the list of predictors. The optical component with the highest correlation among its predictor and the $IA(t)$ of \ac{darm} is considered to be the culprit of the scattered light noise witnessed in \ac{darm}. Visual counterproof can be performed (see figure~6 from~\cite{Longo_2020sc}) overlapping the culprit predictor on the \ac{darm} spectrogram~\cite{Chatterji_2005}.  
The methodology of~\cite{Valdes_2017,Longo_2020sc} was extended and integrated in the \emph{gwadaptive-scattering} pipeline, an automated Python code which allowed to characterize the origin of scattered light glitches in \ac{ligo} during the O3 run~\cite{bianchi2021gwadaptive_scattering}. Furthermore, adaptive analysis can be used to monitor daily the onset and time evolution of scattered light noise in connection with microseismic noise variability ~\cite{Longo:2021avq}. These daily analyses have been integrated in the \emph{gwadaptive-scattering} pipeline as well.


\section{Spectral noise identification and characterization tools}
\label{section:spectral_noise_tools}
\markboth{\thesection. \Sectionname}{}

\subsection{Spectrograms and injected lines identification}
\label{sec:spectro}

Within the \ac{vim} (see section~\ref{sec:tools:VIM}), spectrograms spanning periods from one day to a week are regularly updated using the custom \texttt{Spectro} software~\cite{verkindt_spectro}. This framework is based on a set of ROOT~\cite{BRUN199781,rene_brun_2019_3895860} scripts that provide various indicators (\ac{brms}, Rayleigh spectra, etc.), useful to help investigating non-stationary spectral lines or intermittent noises. The \texttt{Spectro} tool has also been used during O3 to probe the time-frequency pattern of the glitches associated with \ac{bns} range drops.

\subsection{NoEMi and the (known) lines database}
\label{sec:noemi}
The \ac{noemi} tool~\cite{Accadia_2012,noemiwui} tracks on a daily basis spectral lines, both stationary and wandering ones, and searches for coincidences between the lines found in a main channel --- typically the \ac{gw} channel $h(t)$ --- and in a list of auxiliary channels. The \ac{noemi} configuration defines several parameters and thresholds, like for instance: the threshold on the critical ratio\footnote{Defined as the number of standard deviations a given peak amplitude is different from the mean of the peaks amplitude distribution.} for peak selection in the spectra, the frequency resolution (linked to the time length of the data segments over which the \ac{fft} is computed), the name of the main channel, the list of auxiliary channels to search for coincidences. During O2, the \ac{noemi} software produced daily results and looked for peaks in the spectra using a frequency resolution of 1 mHz. With this configuration, \ac{noemi} looked for coincident spectral peaks between the \ac{darm} channel and approximately 40 auxiliary channels.

During the break between the O2 and O3 runs, the \ac{noemi} software has been intensively modified, resolving the main issues identified in the old version. The original 
code was not well-structured (and hence difficult to modify) and also not fully-efficient CPU-wise.
Furthermore, the original version produced several static files which were unessential for the final output. As a further improvement, the MySQL database which stores all parameters of each spectral line found during the run has been normalized, meaning that useless or redundant data have been removed and that the data storage is now more coherent. The database scalability has been improved as well, in order to allow storing more data and handling a higher load of requests. Additionally, a more dynamic interaction with the web interface used to browse the results has been introduced. The new version of the code has been used for the first time in O3.

\begin{figure}
  \center
\includegraphics[width=1\textwidth]{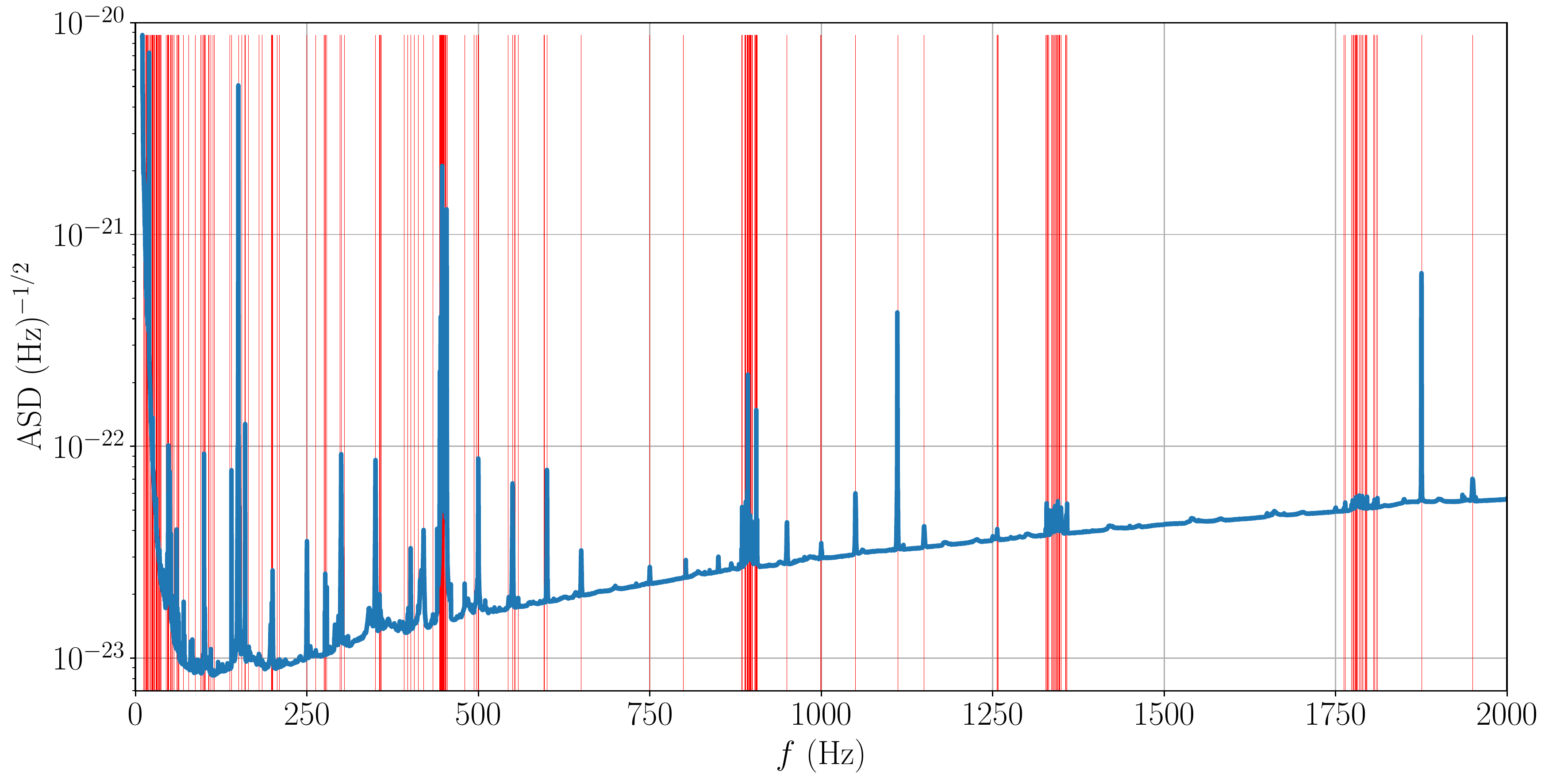}
\caption{Virgo spectral lines identified during O3. The blue curve is the  
estimated \ac{asd} of the Virgo strain channel during O3. Red vertical bars mark the frequency of the identified  
spectral lines. Lines parameters are listed in Reference~\cite{GWOSC_O3_lines}.
Most of the lines have been found by \ac{noemi}.
}
 \label{fig:O3lines_fullband}  
\end{figure}

During O3, \ac{noemi} used the same set of ${\sim}$40 auxiliary channels as in O2, plus an additional set of ${\sim}$140 environmental channels, e.g. seismic, magnetic, and acoustic probes. 
The coincidence between a line in the \ac{gw} strain channel $h(t)$ and the signal of one of the environmental monitor, suggests that the noise line originates from a physical source such as a vacuum pump, a cooling fan, an electronic device, etc. This information helps to identify the instrumental origin of detected lines in $h(t)$, and it has been included in the official Virgo-O3 line list publicly released by the \ac{gwosc}~\cite{GWOSC_O3_lines}. Figure~\ref{fig:O3lines_fullband} illustrates the lines identified in the Virgo \ac{gw} strain channel during O3.

Internally, lines that have been identified are stored in a dedicated database that includes detailed information about them: most notably their times of appearance, and links pointing to the associated documentation (logbook entries, studies, mitigation actions, etc.). The contents of the database can be compared with a new \ac{noemi} processing, to find out quickly which lines identified by \ac{noemi} are already known and which ones are not.

\subsection{Bruco}
\label{sec:bruco}
The \ac{bruco}~\cite{Vajente:2008bka} is a 
python-based tool designed to search for correlated noise among channels.
The code (version 2017-01-23) 
is publicly available from the git repository~\cite{BRUCO_git}, which also provides 
a description of the argument list. An instance 
of the repository is kept with Virgo-specific data
access features.

\ac{bruco} computes the magnitude-squared coherence between a main channel 
(typically, but not necessarily, the detector strain channel $h(t)$)
and all auxiliary channels that, at the time of interest, are recorded by the \ac{daq} system.
Optionally, a set of redundant channels which are known a priori to be correlated with the main one, 
can be excluded. In Virgo, during the O3 run, there were approximately 3,000 non-redundant channels with a 
sampling frequency $\geq$1~kHz.
To deal with the high computational load required by this analysis, \ac{bruco} implements the option of multi-core 
parallel processing in up to $10$ threads.

In the \ac{bruco} implementation adopted for Virgo during O3,
a continuous Science data segment
of length $T=800$~s is selected for the main channel $h(t)$ and, in turn, for each auxiliary channel $n(t)$.
Each data segment is resampled to a targeted output frequency of 2~kHz with the Fourier 
resampler {\it scipy.signal.resample}, 
divided into $N_{ave}=100$ sub-segments (8~s long) and 
the averaged magnitude-squared coherence is computed, as: 

\begin{equation}
C_{h,n}(f_i) = \frac{\mid < FFT_n(f_i)^{*} FFT_h(f_i) >\mid ^{2}}{< \mid FFT_h(f_i)\mid ^{2}>< \mid FFT_n(f_i)\mid ^{2}>}
\label{eq:coherence}
\end{equation}

where $FFT$ denotes the windowed fast Fourier transform, "$<>$" denotes the averaging operation,
$f_{i}$ is the $i^{\textrm{th}}$ frequency bin, and "$^{*}$" the complex conjugate operation.
With these parameters, the frequency resolution is $df = N_{ave} / T = 0.125$~Hz.
Coherence is examined up to 1~kHz, 
and its value is deemed significant if it exceeds a threshold set to 0.03, a value
corresponding to the 95\% confidence level of the distribution of averaged coherence between 
random data~\cite{Piersol}, given the selected parameters.

\ac{bruco} jobs were run regularly and automatically during the whole O3 run.
Daily results are HTML-formatted and made accessible in a dedicated \ac{vim} web page (see 
section~\ref{sec:tools:VIM}). The \ac{bruco} \ac{vim} summary page allows to quickly spot noise paths 
contributing to the \ac{gw} strain channel $h(t)$ in specific frequency bands.
In addition, \ac{bruco} has often been used as an on-demand analysis tool 
to examine specific time periods.

\begin{figure}[htb!]
    \includegraphics[width=0.98\textwidth]{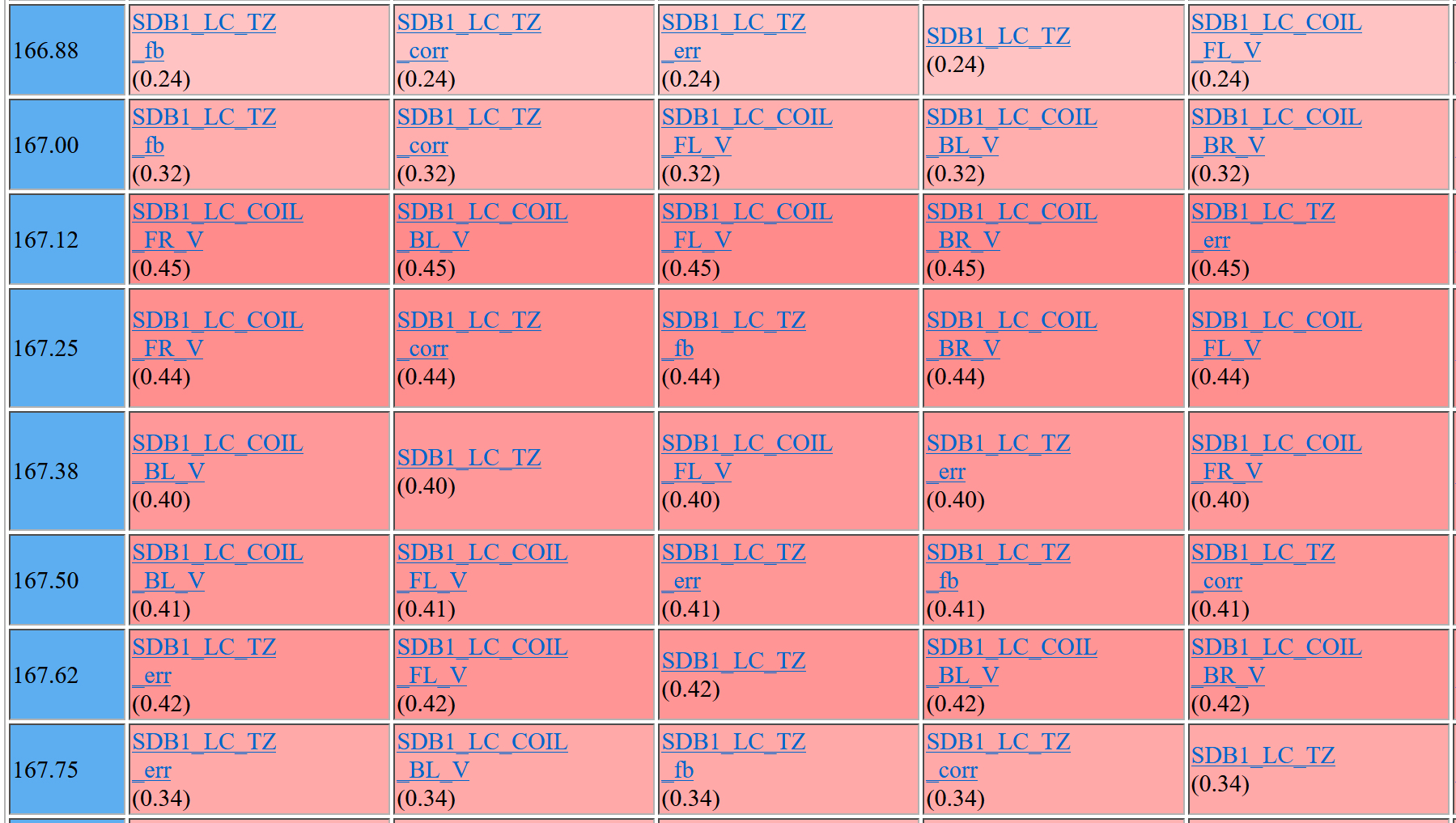}
    \caption{An example of \ac{bruco} result web page displaying, for each frequency bin (leftmost column), the most coherent channels sorted by decreasing coherence value, also represented by the red shade intensity. 
These data are from November 11$^{th}$, 2019.
The large coherence detected at frequencies 155---170~Hz triggered some further investigations of the noise~\cite{Cieslar:elog}.}
    \label{fig:brucopage}
\end{figure}

\begin{figure}[htb!]
    \includegraphics[width=0.9\textwidth]{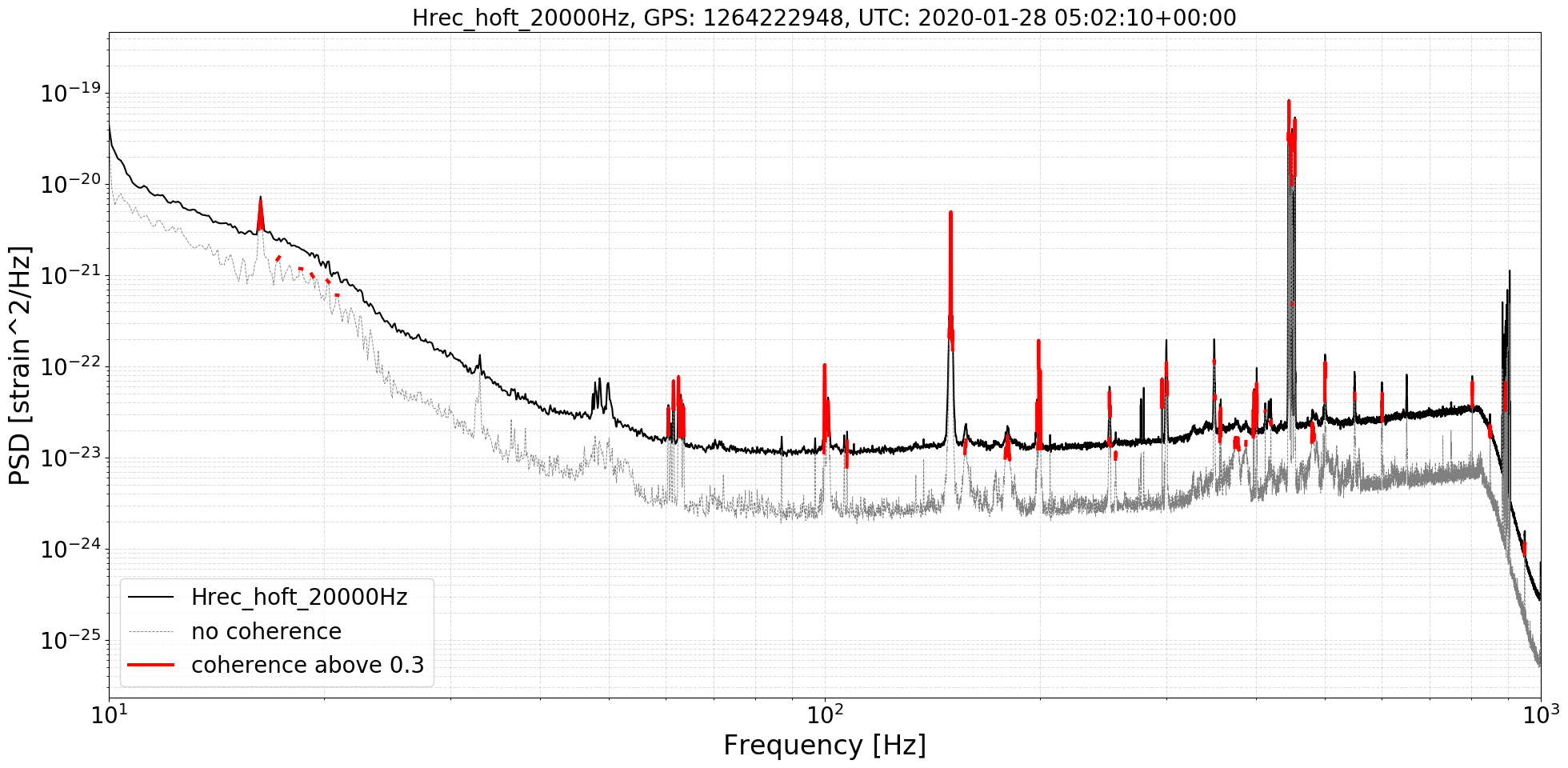}
    \caption{A \ac{bruco} \ac{vim} daily combined projection of the \ac{gw} strain channel $h(t)$ at GPS time 1264222948 (2020/01/28 at 05:02:10 UTC). An anti-aliasing filter with cut-off frequency of 800~Hz is applied to the data. Frequencies where the coherence exceeds 0.3 are highlighted in red, while the dashed grey line shows the noise projection in those regions of the spectrum where the coherence value is below the threshold.}
    \label{fig:bruco_coheprojection}
\end{figure}

\begin{figure}[ht!]
    \centering
	{
	\includegraphics[width=0.98\textwidth]{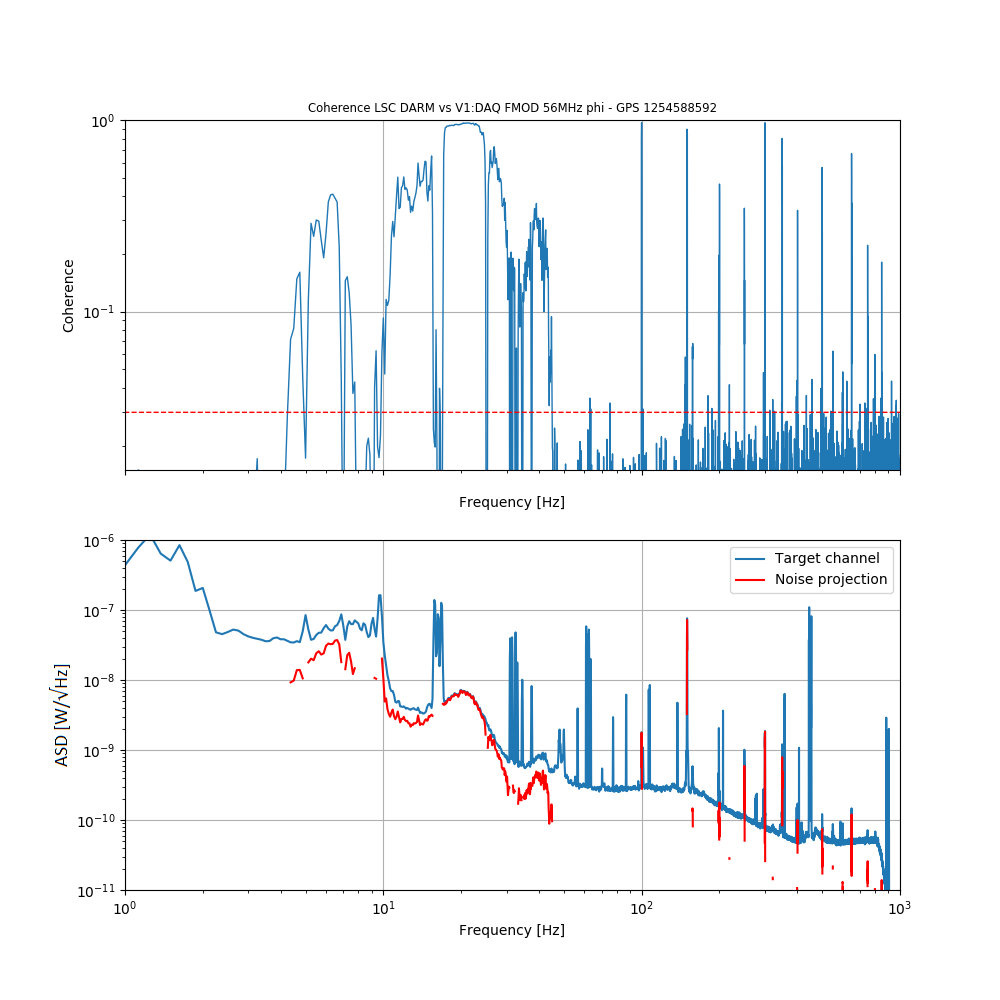}
	}
	\caption{Selection of \ac{bruco} \ac{vim} daily plots evidencing noise contaminating the Virgo \ac{gw} strain channel during O3. The top plot shows the coherence between the \ac{darm} channel and the laser \ac{eom} which produces the 56~MHz signals used for the arms length control.
 The bottom plot shows the \ac{asd} of the \ac{darm} signal (blue line) and the corresponding projected coherence (red line) in the frequency ranges where it was found significant enough. The noise was then found to originate from back-reflected light onto the laser bench, most likely due to a damage on the \ac{eom} for which the component has been removed after O3.}
	\label{fig:bruco_1}
\end{figure}

\ac{bruco} main output is a table that contains, for each frequency bin, the ordered list of the auxiliary channels that are most coherent with the main channel.
The cell background is color-coded in shades of red 
from full red (maximum coherence: 1) to white (no coherence) as shown in figure~\ref{fig:brucopage}.
For each auxiliary channel, a plot (see figure~\ref{fig:bruco_1}) of the {\it projected coherence} 
quantity, $ h_{n}(f)=<FFT_{h}(f)>\sqrt{C_{h,n}(f)}$, is produced and linked to the table. 
In the hypothesis of linear coupling, this quantity estimates the contribution to the strain channel of the noise 
witnessed by the $n^{\textrm{th}}$ auxiliary channel~\cite{Piersol}.
Additionally, the \ac{vim} daily summary page contains the list of the top ranked channels in the frequency 
bins with coherence greater than 0.3 shown in table~\ref{tab:Bruco_cohshort}, and a plot of the combined 
projected coherence greater than 0.5 presented in figure~\ref{fig:bruco_coheprojection}.

\begin{table}[htb!]
    \caption{\label{tab:Bruco_cohshort}An excerpt (15 top lines) of a \ac{bruco} summary table showing the top-ranked channels with coherence greater than 0.3 for GPS time 1264222948 (2020/01/28 at 05:02:10 UTC).
}
    \begin{tabularx}{1.0\linewidth}{r | r | X}
        \toprule
        Frequency [Hz] & Coherence & Type of channel \\
        \hline
  10.9 & 0.329 & Error signal of the second stage of laser frequency stabilization control loop\\
  16.2 & 0.527 & Longitudinal correction applied to the beam splitter payload\\
  17.2 & 0.376 & West arm transmitted light power\\
  18.5 & 0.357 & Same signal\\
  19.2 & 0.342 & Same signal\\
  20.2 & 0.391 & Same signal\\
  20.8 & 0.318 & Same signal\\
  21.8 & 0.327 & Same signal\\
  33.0 & 0.417 & Error signal of the second stage of laser frequency stabilization control loop\\
  60.4 & 0.579 & Calibration signal applied to the \ac{we} test mass \\
  61.0 & 0.444 & Signal of the phase camera on the external detection bench\\
  61.4 & 0.817 & Calibration signal applied to the \ac{bs} payload\\
  61.5 & 0.944 & Calibration signal applied to the \ac{ne} test mass\\
  62.500 & 0.929 & Calibration signal applied to the \ac{we} test mass\\
  99.800 & 0.356 & Experimental power grid voltage monitor signal\\
        \bottomrule
    \end{tabularx}    
\end{table}

Figure~\ref{fig:bruco_1} shows \ac{bruco} daily plots illustrating one example of noise contamination spotted during O3, which triggered a more in-depth investigation~\cite{EnvHuntVirgoO3,Was2:elog}.

\subsection{MONET}
\label{sec:monet}
The interferometer noise spectrum can sometimes present peculiar structures as a consequence of non-linear couplings between different noise processes. Some of these structures consist of two pairs of sidebands around known spectral lines, which are not explained by means of the previously described linear coherence methods. One example of this kind of noise is the bilinear noise, generated by the coupling of two noise sources that jointly affect a third signal. In \ac{gw} detectors, the main cause of this bilinear noise is the  up-conversion of the low frequency seismic noise, that can affect the mirrors angular controls, which couples with some narrow-band noise processes, like power mains lines ($50~\mathrm{Hz}$ fundamental frequency or its harmonics) or calibration lines (see for example~\cite{DiRenzo:2020,Aasi:2012wd}).

The \ac{monet}~\cite{MONET-tds} is a python-coded framework designed to investigate these sidebands. The main hypothesis at the basis of this tool is that the sidebands are due to the coupling of a carrier signal with the low-frequency (with a typical cutoff frequency $f_{c}$ of a few~Hz) part of an auxiliary channel, called the modulator signal. Under this hypothesis, \ac{monet} searches for coherence between a main channel (typically, but not necessarily, the detector strain channel or the \ac{darm} channel) and a new signal given by  the product, in the time domain, of the chosen carrier signal and a modulator signal.
That carrier signal can be a real channel or a simulated signal such as, for example, a sinusoidal signal.

Similarly to what is done with \ac{bruco},
continuous Science data segments of a specific time length $T$ are selected for all the channels to be investigated with \ac{monet}. Then, each data segment is resampled to a targeted output frequency (\mbox{$f^{\rm out}$}) and, finally, the magnitude-squared coherence is computed using equation~\eqref{eq:coherence}. For the analysis of the O3 Virgo data, we typically used the following values: $f_{c}=5~\mathrm{Hz}$, $T=1200~\mathrm{s}$ and $f^{\rm out}=1~\mathrm{kHz}$.

\ac{monet} can be executed on demand, investigating dozens of auxiliary channels and spectral lines in every single processing. The outputs are made available through a hierarchical structure:

\begin{itemize}
\item[$\bullet$] A main directory, whose name is built from the main channel name, the initial GPS time and the time length of the segment of data to be analysed.
\item[$\bullet$] The main directory contains one sub-directory for each carrier signal used. 
\item[$\bullet$] Each sub-directory contains in turn several folders, one for each modulator channel.
\end{itemize}

\begin{figure}[h!]
\begin{center}
\includegraphics[width=0.9\textwidth]{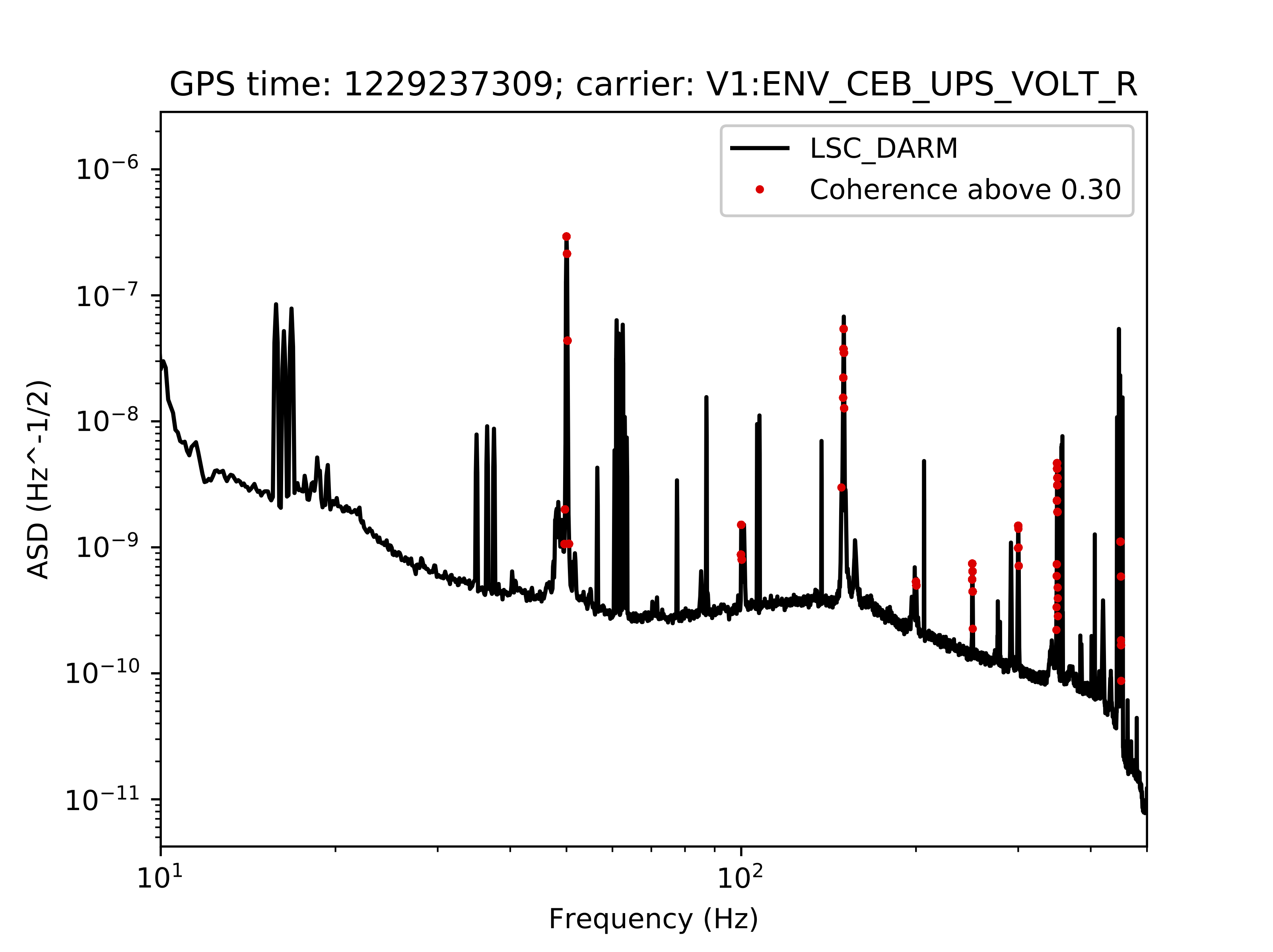}
\caption{\ac{asd} of the \ac{darm} channel (in black), together with red points that mark the frequencies at which the coherence found by \ac{monet} is above the threshold, fixed in this case at $0.3$. The initial GPS time of the analysed data and the chosen carrier signal (a voltage monitor of the uninterruptible power supply in the \ac{ceb}) are indicated on the top of the figure.} \label{fig:ASD-monet}
\end{center}
\end{figure}

The sub-directory contains a table in text format and a summary figure. The table (see table~\ref{tab:Table-monet} for an excerpt example) is made of three columns and several rows. Each row corresponds to a different frequency bin, indicated in the first column. The second and third columns display, for each modulator channel, the computed above-threshold coherence value and the name of that channel respectively.
Figure~\ref{fig:ASD-monet} shows a \ac{monet} summary plot for the \ac{darm} channel during the pre-O3 commissioning phase.
Red points marking the frequencies at which coherence above threshold is found with at least one auxiliary channel are superimposed to the main channel \ac{asd}.  
 For instance, the red points at ${\sim} 200$~Hz shown in figure~\ref{fig:ASD-monet} correspond to the frequency bins reported in table~\ref{tab:Table-monet}; at such frequencies, a coherence value above the chosen threshold was found for several modulator channels. 

\begin{table}
    \caption{An excerpt from a \ac{monet} table with coherence values above the threshold, fixed in this case at 0.3. The initial GPS time of the analysed data is 1229237309 (2018/12/19 at 06:48:11 UTC). Main channel: the \ac{darm} one; secondary channel: a voltage monitor of the uninterruptible power supply in the \ac{ceb}.}\label{tab:Table-monet}
    \begin{tabularx}{1.0\linewidth}{r | r | X}
        \toprule
        Frequency [Hz] & Coherence & Channels \\
       \hline
	200.0 & 0.313---0.347 & Some control signals, mirror longitudinal corrections (applied by varying the current flowing into the coil-magnet actuators for the corresponding test masses), and DC laser powers measured in various locations of the interferometer. \\ 
       \hline
	200.2 & 0.379---0.437 & Same channels as above. \\
       \hline
	200.3 & 0.456---0.496 & Same channels as above. \\     
       \bottomrule
    \end{tabularx}
\end{table}

\begin{figure}[h!]
\begin{center}
\includegraphics[width=0.9\textwidth]{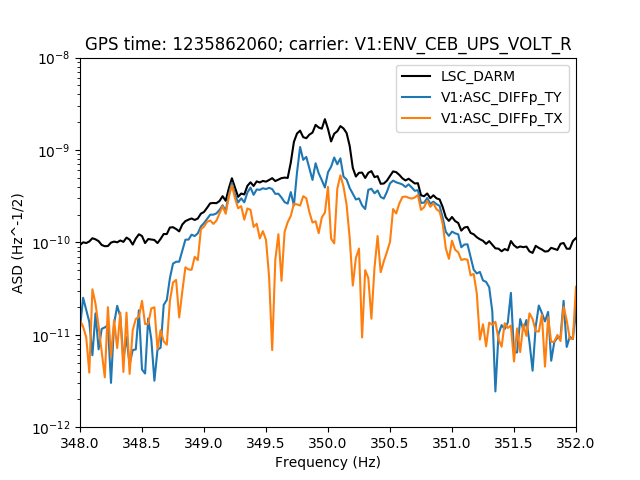}
\caption{\ac{asd} of the \ac{darm} channel around the 350~Hz spectral line (in black), together with the noise projection, based on the \ac{monet} coherence values obtained with the modulator channels \texttt{ASC$\_$Diffp$\_$TY} and \texttt{ASC$\_$Diffp$\_$TX}, in blue and orange, respectively. These two channels are used to control the angular motion of the mirrors in the detector arms, in order to guarantee the proper recombination of the laser beams at the beam splitter. The initial GPS time of the analysed data and the chosen carrier signal are indicated on the top of the figure.}\label{fig:ASDzoomed-monet}
\end{center}
\end{figure}

In each innermost folder, a table and a plot are generated, in which the coherence values associated with the specific modulator channel for each frequency bin are reported. Several other plots are also produced, in which the \ac{asd} of the main channel is reported, together with the noise projection based on the coherence values, around the specific spectral lines to be investigated as  seen in figure~\ref{fig:ASDzoomed-monet}.

The results of \ac{monet} can help identify the origin of the observed sidebands in the main channel.
The modulating channels that, once multiplied by the carrier signal, exhibit the largest coherence with the main channel can witness those parts of the detector that are more sensitive to low frequency noise, usually due to microseismic activity, and its up-conversion through bilinear couplings.

During O3 and the preceding commissioning phase~\cite{fiori-monet:elog,fiori2-monet:elog,fiori3-monet:elog}, \ac{monet} has been effectively used to characterize the spectral noise in the strain channel and guide the commissioning activity on how to improve it.
This has been the case for the observed sidebands of the $150 ~\mathrm{Hz}$ mains frequency harmonic line~\cite{direnzo-monet:elog}.
The increase in the noise around this frequency during intense microseismic activity has been found coherent with the angular controls of the \ac{bs} as modulating channels. 
Moreover, under similar noise conditions, the value of the coherence, hence the strength of the bilinear coupling and the observed overall noise level at $150~\mathrm{Hz}$, was proven to depend on the way the \ac{bs} is controlled (``full bandwidth'' vs.~``drift control''), convincing the team working on the interferometer alignment to prefer to switch the operational mode of the \ac{bs} during high microseismic activity with an improvement of up to $2~\mathrm{Mpc}$ of \ac{bns} range~\cite{DiRenzo:2020}.

In addition to that, \ac{monet} has helped characterize the effect of the beam alignment on the \ac{pr} mirror on the linear noise subtraction of the frequency noise coupling due to the arm asymmetry used to produce the reconstructed \ac{gw} strain signal $h(t)$.
Large values of coherence are found with the channels controlling the \ac{pr} alignment at $1111~\mathrm{Hz}$, the frequency of the injected line for the laser frequency stabilization control loop, when the efficiency of the noise subtraction was smaller~\cite{mwas-prcl:elog}. 
This has been interpreted as a consequence of the \ac{pr} mirror transverse position modulating the frequency noise coupling, hence producing non-linear effects, not adequately taken into account by means of the linear subtraction method~\cite{mwas-monet:elog}.
This effect has been quantified in a reduction of $2~\mathrm{Mpc}$ of \ac{bns} range when the beam is transversely misaligned on the \ac{pr}.


\section{Common LIGO-Virgo tools}
\label{section:common_tools}
\markboth{\thesection. \Sectionname}{}

\subsection{DQSEGDB}
For each data quality flag, the \ac{dqsegdb}~\cite{dqsegdb} stores the segments (integer GPS ranges) during which that particular flag is active, meaning that the set of conditions it is based on is fulfilled. For instance, one such flag tags the GPS segments during which the Virgo detector is taking data in Science mode.
 There are two ways to fill this database with Virgo flags:

\begin{itemize}
\item Online, during the data taking, through the \texttt{SegOnline} server~\cite{O3DetChar_results} that is compiling information provided by various data streams;
\item Offline, by completing or fixing existing segment sets, or adding new data quality flags to monitor additional conditions.
\end{itemize}

A versioning system is used to keep track of changes in segment lists that can modify a particular flag, i.e. that impact offline analyses, by changing the contents of the dataset they are processing. By convention, the highest version number corresponds to the best (most recent) segment list and is the one queried by default.

\subsection{GraceDB}
During O3, the \ac{gracedb}~\cite{gracedb} has been the central place where information about transient \ac{gw} candidates was uploaded and stored: online search triggers, source localization estimates in the sky, data quality information, other metadata, etc. In particular, \ac{gracedb} triggered automatically frameworks like the \ac{dqr} through the \ac{lvalert}~\cite{lvalert} when candidate events of interest were identified; and, consequently, \ac{dqr} results (see section~\ref{subsection:DQR}) got uploaded back to \ac{gracedb} as soon as they became available. \ac{gracedb} has a public-faced portal that provides information about the public alerts shared with the astronomer community, while most of its data are private and reserved to the \ac{ligo}, Virgo and \ac{kagra} collaborations.


\section{Data Quality Reports}
\label{subsection:DQR}
\markboth{\thesection. \Sectionname}{}

\subsection{Introduction}

The \ac{dqr} is a framework developed by \ac{ligo} and Virgo for the O3 run, in order to quickly gather enough information to vet the significant triggers found by the online transient \ac{gw} searches. The goal is to either confirm the associated public alert, or have it retracted at once. All 80 public alerts delivered during O3~\cite{public_alerts} (of which 24 have been retracted) have used this input.

A \ac{dqr} runs on a computing cluster where the $h(t)$ strain channel and the associated raw data auxiliary channels are available in low latency. Therefore, each collaboration (Virgo at \ac{ego} and \ac{ligo} with its two detectors) was responsible for the implementation, the operation, the monitoring and the upgrade of its own \ac{dqr} framework. There was however an agreement on a common format for the check outputs, originally developed by \ac{ligo}~\cite{dqr}.

The \ac{dqr} framework is triggered by \ac{gracedb} through the \ac{lvalert} protocol. A \texttt{JSON} payload received from \ac{gracedb} allows for the generation and the configuration of a new \ac{dqr}. The checks are then processed and their results are uploaded back to \ac{gracedb}, alongside all the records associated with that particular \ac{gw} candidate.
In order for the \ac{dqr} to be triggered, a candidate event is required to have a false-alarm rate below 1/day.
This is a conservative threshold, much higher than that required to release the candidate as a public alert, but still low enough to keep the computational cost of generating the \acp{dqr} under control. 
Therefore, in average, only a handful of \acp{dqr} were automatically processed on a daily basis during the ${\sim}330$~days of the O3 run: not a high CPU load overall, but still about 20 times more \acp{dqr} than the number of public alerts that had to be vetted.

\subsection{Virgo implementation and contents}

\begin{figure}
  \center
  \includegraphics[width=\textwidth]{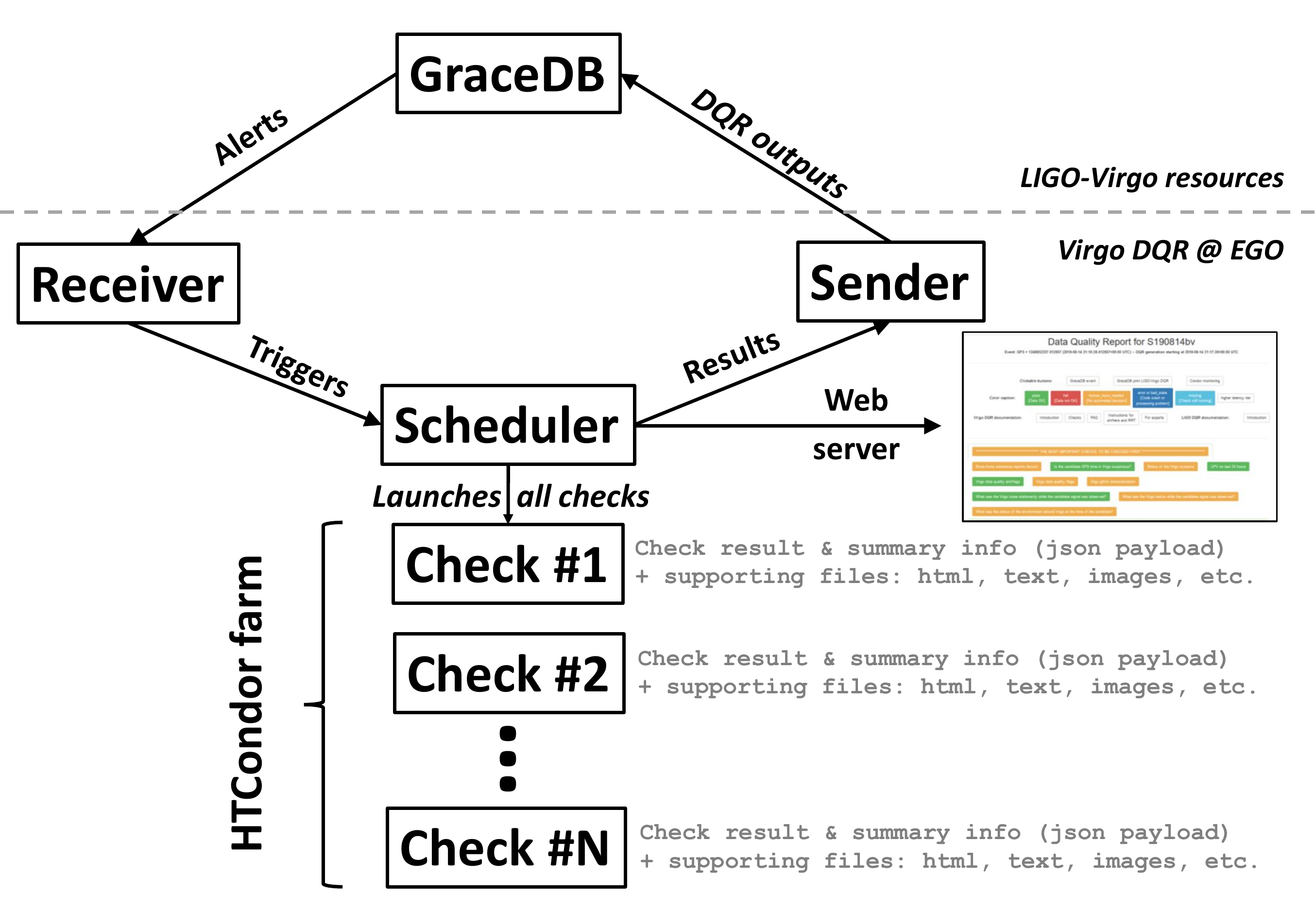}
  \caption{Schematics of the Virgo O3 \ac{dqr} architecture. See text for a description of this workflow.}
  \label{fig:DQR_architecture}  
\end{figure}

Figure~\ref{fig:DQR_architecture} summarizes the Virgo \ac{dqr} architecture used during the O3 run. When a trigger with a low-enough false alarm rate is received, a new \ac{dqr} is created and configured, using information from \ac{gracedb}. Then, the data quality checks are run in parallel on the \ac{ego} HTCondor~\cite{condor-hunter} farm. As soon as a given check is complete, its results are uploaded back to \ac{gracedb}. In parallel, the \ac{dqr} progress and results are immediately available for Virgo DetChar experts and on-duty people, through an \ac{ego}-internal web server. The \ac{dqr} format~\cite{dqr}, originally developed by \ac{ligo}, is lighter and more versatile than the \ac{gracedb} user interface: it ensures a direct access to the Virgo \ac{dqr} outputs. The \ac{dqr} web page URL is automatically sent to the relevant internal mailing list as soon as the newly-created \ac{dqr} processing starts.

The Virgo \ac{dqr} framework has evolved quite significantly over the course of O3. Partly to tune and improve the workflow based on the experience accumulated when stressing the system during the actual run, but mainly to extend the scope of the \ac{dqr} by adding additional data quality checks. These new checks were either tests that had been foreseen but could not have been implemented by the start of O3, or new procedures that brought additional information that was found missing or useful when vetting real O3 triggers.

Therefore, at the end of O3, the Virgo \ac{dqr} included 34 checks, for a total of 99 jobs. There are roughly three jobs per check: the first, to run the code and process the data; the second, to post-process the check results and convert them to the \ac{ligo}-Virgo common \ac{dqr} format; finally, the third to upload the results back to \ac{gracedb}. Some checks included a fourth job as an initial configuration phase while others, developed specifically for the \ac{dqr}, produced directly check outputs in the required \ac{dqr} output format, meaning that those checks required one job less.

The Virgo \ac{dqr} checks have been categorized in the following way:

\begin{itemize}
\item Key checks \\ They bring information mandatory to properly vet a candidate event. This includes: the top-level status of the detector at the time of the trigger; some time-frequency spectrograms of the \ac{gw} strain data at different timescales around that time; finally, the scan of the main data quality flags available online, in order to look for any obvious problem in the data.
\item Characterization of the Virgo detector noise around the time of the trigger \\ The noise transients (glitches) are inventoried and their potential overlap with the time-frequency extent expected for the candidate is probed, if applicable. In addition, searches for noise correlations in the time domain and noise coherences in the frequency domain are run, such as tests of noise Gaussianity and stationarity.
\item Detailed Virgo status \\ Several different analysis contribute to this global picture of the instrument. All data quality flags available are checked. In addition, the \ac{dms} database is scanned to extract the snapshots closest in time to the trigger, to see what warnings or alarms were on, if any. Also, the logfiles of all the online servers running in the \ac{daq} are scanned to spot errors that could be coincident with the trigger or impact it. Finally, various live data/reference comparison plots are generated to check the time series and distributions of a subset of the \ac{daq} channels.
\item Digest of the environment status \\ This includes checking the seismic noise at \ac{ego} in various frequency bands corresponding to different sources (microseism related to sea activity on the Tuscany shoreline or local anthropogenic activities: see~\cite{o3virgoenv} for details), the sea activity and the weather.
\end{itemize}


\section{Conclusion}
\label{section:outlook}
\markboth{\thesection. \Sectionname}{}
This article has presented the diverse tools used by the Virgo DetChar group to fulfill its tasks during the O3 run. These tools will again form the basis of the DetChar framework for the next \ac{ligo}-Virgo-\ac{kagra} observing run O4, currently scheduled to start in Spring 2023. As the sensitivity of the detectors is expected to be improved, a larger rate of detections should follow. To prepare this new challenge, work has been ongoing for the past two years and will continue in the coming months. Improvements are manifold: first on the core software itself, by improving the performance of the existing packages and extending their functionalities. A priority is to automate analyses further and to have them run more often, so that results are readily available and simpler to use. Another avenue being explored is the use on Virgo data of tools which have been developed by our \ac{ligo} colleagues: \texttt{iDQ}~\cite{Essick_2020} to help identifying glitches in the \ac{gw} strain channel $h(t)$ or \texttt{STAMP-PEM}~\cite{STAMP-PEM}, a framework similar to \ac{bruco} but designed to run on much longer stretches of data and useful for the analyses searching for continuous \ac{gw} signals. Finally, new tools are being developed as well, one example being the hunt for scattering light-induced noise, one of the main limitations of the current ground-based \ac{gw} detectors. These additional analyses will require dedicated computing resources which will be provided by the Virgo Tier1 computing centers, in particular the CC-IN2P3 in France.


\section*{Acknowledgments}
The authors gratefully acknowledge the Italian Istituto Nazionale di Fisica
Nucleare (INFN), the French Centre National de la Recherche Scientifique (CNRS)
and the Netherlands Organization for Scientific Research (NWO), for the construction
and operation of the Virgo detector and the creation and support of
the EGO consortium. The authors also gratefully acknowledge research support
from these agencies as well as by the Spanish Agencia Estatal de Investigaci\'on,
the Consellera d'Innovaci\'o, Universitats, Ci\`encia i Societat Digital de la
Generalitat Valenciana and the CERCA Programme Generalitat de Catalunya,
Spain, the National Science Centre of Poland and the European Union --- European
Regional Development Fund; Foundation for Polish Science (FNP), the
Hungarian Scientific Research Fund (OTKA), the French Lyon Institute of Origins
(LIO), the Belgian Fonds de la Recherche Scientifique (FRS-FNRS), Actions
de Recherche Concert\'ees (ARC) and Fonds Wetenschappelijk Onderzoek --- 
 Vlaanderen (FWO), Belgium, the European Commission. The authors gratefully
acknowledge the support of the NSF, STFC, INFN, CNRS and Nikhef for
provision of computational resources.

{\it We would like to thank all of the essential workers who put their health at risk
during the Covid-19 pandemic, without whom we would not have been able to
complete this work.}

{\it The authors would also like to thank Samuel Salvador for his extensive and careful proofreading of the manuscript.}

\markboth{\Sectionname}{}


\addcontentsline{toc}{section}{List of Abbreviations}
\printacronyms[name=List of Abbreviations]


\clearpage

\section*{References}
\markboth{REFERENCES}{}
\bibliographystyle{iopart-num-doi}
\bibliography{references}

\end{document}